\documentclass[flushrt,preprint2]{aastex}

\newcommand{\apl}{\lesssim}

\newcommand{\etal}{{\it et al.\ }}
\newcommand{\eg}{{\it e.g.\ }}
\newcommand{\ie}{{\it i.e.\ }}
\newcommand{\kms}{~{\rm km\ s^{-1}}}

\slugcomment{Submitted to \apj}
\shortauthors{Fern\'andez-Soto et al.} 
\shorttitle{A Blind Test of Spectroscopic Redshifts}

\begin{document}

\title{On the compared accuracy and reliability of spectroscopic and
photometric redshift measurements}

\author{Alberto Fern\'andez-Soto\altaffilmark{1,2,3}, Kenneth M. 
Lanzetta\altaffilmark{1},  Hsiao-Wen Chen\altaffilmark{4},\\ Sebastian M. 
Pascarelle\altaffilmark{1}, and Noriaki Yahata\altaffilmark{1}}
\altaffiltext{1}{Department of Physics and Astronomy, State University of New 
York at Stony Brook,\\ Stony Brook, NY 11794-3800, U.S.A. (fsoto, lanzetta, 
sam, nyahata@sbastr.ess.sunysb.edu)}
\altaffiltext{2}{Osservatorio Astronomico di Brera, Via Bianchi 46, Merate
(LC), I-23807, ITALY}
\altaffiltext{3}{Marie Curie Fellow}
\altaffiltext{4}{The Observatories of the Carnegie Institution of Washington,
813 Santa Barbara Street, Pasadena, CA 91101, U.S.A. (hchen@ociw.edu)}

\begin{abstract}
  We present a comparison between the catalog of spectroscopic redshifts in
the Hubble Deep Field (HDF) recently published by Cohen and collaborators
and the redshifts that our group has measured for the same objects using
photometric techniques. This comparison is performed in order to fully
characterize the errors associated to the photometric redshift technique. The
compilation of spectroscopic redshifts incorporates previously published
results, corrections to previously published wrong values, and new data, and
includes over 140 objects in the HDF proper. It represents the deepest,
cleanest, most complete spectroscopic catalog ever compiled. We particularly
study each and every object for which our redshift and the one measured by
Cohen and collaborators seem to disagree. In most of those cases the
photometric evidence we put forth is strong enough to call for a careful
review of the spectroscopic values, as the spectroscopic values seem to be in
error. We show that it is possible to characterize the systematic errors
associated to our technique, which when combined with the well-measured
photometric errors allow us to obtain complete information on the redshift of
each galaxy and its associated confidence interval, regardless of its
apparent magnitude.  One of the main conclusions of this study is that, to
date, all the redshifts from our published catalog that have been checked
have been shown to be correct (within the stated confidence limits).  This
implies that our set of spectrophotometric galaxy templates is a fair
representation of the galaxy population at all redshifts ($0<z<6$) and
magnitudes ($R<24$) explored to date. On the other hand, spectroscopy of
faint sources is subject to unknown and uncharacterized systematic
errors. These errors will in turn be transmitted to any photometric redshift
technique which uses spectroscopic samples in its calibration. Our analysis
proves that photometric redshift techniques can and must be used to extend
the range of applicability (in redshift, signal-to-noise, and apparent
magnitude) of the spectroscopic redshift measurements.
\end{abstract}


\clearpage

\section{INTRODUCTION}

  Measuring redshifts of extragalactic objects through spectroscopy is a hard
task. This is particularly true in those extreme cases when the available
technology is strained to its limits and applied to objects which are faint
and consequently offer low signal-to-noise spectra. Even in cases where a
good signal-to-noise ratio is achieved, the absence of clear and
characteristic features in the observed spectral range can hamper the
observer's ability to determine a trustworthy redshift. Moreover, the process
of line identification which is necessary in order to measure a redshift is
very often too subjective, and hence likely to be biased. In order to avoid
this problem, various techniques have been developed that try to automate the
process, amongst them the application of spectral cross-correlation,
principal component analysis, and combinations thereof---see for example
\cite{Glazebrook98}. However, these methods still suffer from numerous
problems produced by low signal-to-noise spectra, instrument-related
systematics, or inability to handle samples covering large redshift ranges.

  A completely orthogonal approach consists in the use of purely photometric
information on the objects under study. In this approach, the weight of the
redshift identification is carried by the broad-band continuum shape and the
presence/absence of spectral breaks, such as the one at 4000 \AA, or the
onset of the Lyman $\alpha$ forest and the Lyman limit at high redshifts,
rather than by narrow-band emission/absorption features. Photometric redshift
methods like these have been known and used for many years---for example by
\cite{Butchins81}, \cite{Loh86}, or \cite{Connolly95}---but the interest in
them has grown enormously since the Hubble Space Telescope (HST) produced the
Hubble Deep Field (HDF) observations \citep{Williams96}. The reason for this
interest is that the HDF observations are ideally suited for the application
of photometric redshift techniques: they include multi-band information
spanning all of the visible and near-IR range and they are extremely deep
[reaching to AB(8140) $\approx$ 29]. This second fact also implies that it is
utterly impossible to obtain spectra of the bulk of the detected objects, and
given the extremely high density of galaxies it is even difficult sometimes
to measure the spectra of the (relatively) bright ones.

  Since the data became available in December 1995, several papers have been
published giving spectroscopic redshifts of over a hundred objects in the HDF
[see references in \cite{C00}, hereafter C00]. Several groups have also
published the results of different photometric redshift techniques applied to
the field, with catalogs of photometry and photometric redshifts:
\cite{LYF96} (LYF96 hereafter), \cite{Gwyn96}, \cite{Sawicki97},
\cite{Wang98}, \cite{FLY99} (FLY99 hereafter), \cite{Furusawa00},
\cite{Benitez00}. At least one comparison of photometric and spectroscopic
redshifts has been published \citep{Hogg98}, with ``photometrists'' supplying
answers to the questions of the ``spectroscopists,'' who compared the
former's results with the ones measured from their spectra.  This comparison
showed that photometric techniques, as a whole, can give relatively accurate
measurements (to within $\Delta z/(1+z) \approx 0.10$ rms), and that they are
very well suited for the analysis of large samples of deep, multi-color
imaging data.

  However, it is not so well known that photometric redshift techniques have
also been instrumental in detecting some incorrect spectroscopic
redshifts. Our group made public as early as 1997 a list of ``successes'' and
``failures'' of our photometric redshift technique, in which it was shown
that many of the discordances observed in the HDF $z_{\rm phot}$ vs $z_{\rm
spec}$ plots could actually be due to mistakes in the spectroscopic
redshifts---and some of them certainly were, as acknowledged by the
respective observers afterwards [see \cite{Lanzetta98} for
details]. Apparently, it has not been sufficiently remarked that detecting
and identifying the narrow features which are necessary to unambiguously
determine a spectroscopic redshift is a very hard task in the low S/N spectra
we are confronting.

  Another proof of the quality of the photometric redshifts is the fact that
a number of objects for which our own group measured redshifts $z_{\rm phot}
> 5$ have been observed with the Keck telescope, and confirmed to be amongst
the most distant objects ever observed. For example, \cite{Spinrad98}
measured $z_{\rm spec}=5.34$ for object \#3 in FLY99 ($z_{\rm phot}=5.28$),
and \cite{Weymann98} measured $z_{\rm spec}=5.60$ for object \#213 in the
same catalog ($z_{\rm phot}=5.64$).

  Recently, Cohen and collaborators have published a complete list of
spectroscopic redshifts in the HDF and flanking fields (C00). They have
compiled all the data published to date, eliminated and/or ammended the
redshifts that have been shown to be in error, and added a number of
observations of their own. Their final list includes 671 entries, 158 of
which are within the HDF proper (we have moved nine objects which are located
in the PC from their ``flanking fields'' table to the HDF table). After
excluding 11 stars and 1 object (HDF36774\_1235) which does not lie inside
the area we studied, we are left with 146 galaxies using which both methods
are compared.

  In this paper, the catalog of spectroscopic redshifts compiled by
\cite{C00} is compared to the catalog of photometric redshifts obtained by
our group. Cohen's catalog is the deepest (reaching down to $R=24$ and
beyond), most complete (92\% complete to that limit) and cleanest (at least
seven redshifts have been corrected from previously published values which
were shown to be in error) ever published. By performing this comparison on a
sample of bright objects---where the photometric uncertainties are small---we
expect to obtain a complete characterization of the systematic errors
associated with our technique, and hence to be able to measure accurate
errors associated to any of our photometric redshift determinations. A first
comparison shows that over 90\% of the residuals are well represented by a
gaussian distribution with an rms dispersion $\sigma_{\Delta z/(1+z)} \apl
0.08$ at all redshifts. Nine objects out of 146 have residuals which are more
than $4\sigma_{\Delta z/(1+z)}$ away from the spectroscopic value. A very
interesting result comes from the detailed analysis of these nine objects
with discordant redshifts: we show in this paper that {\it at least} five of
the discordant points can be due to errors in the {\it spectroscopic} values,
with the proportion being possibly as high as 90\% (eight out of nine). In
the only case in which the photometric redshift technique is responsible for
the discordance, we show that when the full redshift likelihood function and
the systematic errors---as estimated from the whole sample---are taken into
account, the spectroscopic redshift is in fact {\it well within $2\sigma$ of
the photometric redshift}.

  These results lead us to conclude that the spectroscopic redshift technique
is plagued by unknown and uncharacterized errors, which become problematic,
or even dominant, when low signal-to-noise spectra are analysed. These errors
are then transmitted to those photometric redshift techniques that make use
of spectroscopic redshifts as ``calibrators.'' On the other hand, the
photometric redshift technique we have presented in FLY99 has {\it as of
today} not been shown to be in error over the stated confidence limits in a
single case, and has the advantage of including a full characterization of
its errors, through the use of the redshift likelihood function.

  Our experience shows that, as a general rule, the systematic errors
dominate our technique for objects down to $m=m_{\rm lim}-2$, where $m_{\rm
lim}$ is the detection limit in the images. This means that, for HDF-style
surveys, redshifts can be obtained with $\sigma_{\Delta z/(1+z)} \apl 0.08$
for objects down to $AB \approx 26$, two magnitudes below the limits
attainable by traditional spectroscopy with the largest telescopes. For
magnitudes fainter than this limit, the photometric errors dominate over the
systematic ones, and the errors associated with the photometric redshift
become larger. In both cases the accuracy of the method should remain
constant, as long as our set of template spectral energy distributions (SEDs)
remains a fair representation of the observed galaxy population. This has
been proven to be the case at all redshifts ($0<z<6$) and magnitudes ($R<24$)
probed to date. This paper further proves that there is a strict upper limit
(less than 5.4\%, to $2\sigma$ confidence) to the possible percentage of
galaxies whose SEDs differ significantly (within the available photometry)
from our templates in a flux limited ($R<24$) sample.

  The structure of this paper is as follows: In \S 2 we present the
data. Section 3 introduces the comparison, and in \S 4 we fix our attention
on the objects for which the spectroscopic and photometric redshifts
disagree. Section 5 presents the data on one particular object which does not
show up as ``discordant'' but for which we are led to believe the
spectroscopic redshift is in error.  In \S 6 we discuss our results and
in \S 7 we enumerate our conclusions.

\section{THE COMPARISON REDSHIFT CATALOG}

  In this Section we present the photometric and spectroscopic catalogs 
used in the comparison, and the process of cross-identification of the 
objects in both lists.

\subsection{The Spectroscopic Redshift Catalog}

  The sample of spectroscopic redshifts is the one given in C00. It contains
671 entries and is, as stated by the authors, complete to over 92\% for
objects inside the HDF area to $R=24$, irrespective of morphology. Some
fainter objects are also included. All spectroscopy involved was performed
using LRIS on the Keck Telescopes. It is very important to note that this is
{\it the cleanest, deepest, most complete} list of spectroscopic redshifts
available for the HDF, which is to say, the cleanest, deepest, most complete
sample of spectroscopic redshifts available to date.

  The authors of C00 also remark that all redshifts listed have been
carefully checked. In this process several values have been changed from
previously published papers: six objects from \cite{Cohen96}, two objects
from \cite{Steidel96}, two objects from \cite{Phillips97}, and an unspecified
number of objects from \cite{Hogg98} and the Hawaii web database\footnotemark
\footnotetext{{\tt http://www.ifa.hawaii.edu/$\sim$cowie/tts/tts.html}}. As
explained in \S 1, some of these changes were actually induced by comparison
of the original spectroscopic values and the photometric redshifts as
supplied by ourselves and other groups. The final list includes 146 objects
with reliable spectroscopic measurements: the 149 listed in Table 2b of C00,
plus nine objets from their Table 2a---Flanking Fields---that lie within the
PC and, hence, enter our analysis, minus 11 stars, and minus one object
(HDF36377\_1235) which does not lie in the area studied by us (see footnote
about this object in Section 2.3).

\subsection{The Photometric Redshift Catalog}

  Our photometric redshift sample is basically the one published in FLY99.
We used the photometric catalog as described in that paper, but we applied to
it the more accurate photometric redshift technique described in \cite{Y00}
(Y00 hereafter). The main difference with respect to the FLY99 paper is the
inclusion of two starburst spectrophotometric templates, which improves the
behavior of our method in the redshift range $z=2.0-4.0$ (where our former
method systematically underestimated the redshifts by $\Delta z \approx
0.5$), as demonstrated by \cite{Benitez00}. We have also decreased the
calculation step from $\delta z=0.04$ to $\delta z=0.01$, to take full profit
of the excellent quality of the measurements. 

  It must be remarked that {\it the objective of our technique is not to fit
the observed photometry}, but to obtain a best-fit measurement of the
redshift. It is important to notice that in the case of bright objects our
fit will never be a ``perfect fit'' (the photometry being too accurate for
our sample of SEDs to reproduce every possible spectrum) whereas, in the
opposite end, for very faint objects any basically correct SED will represent
an equally good fit to the observations. Hence, the particular value of the
maximum likelihood (or the value of the minimum $\chi^2$) cannot be taken as
a measurement of the goodness of the estimated redshift.

  The sample includes 1067 galaxies, with photometric redshifts spanning the
range $z \approx 0-6$, and is complete to $AB(8140) = 26.0$ over 5.2 square
arcminutes and to $AB(8140)=28.0$ over 3.9 square arcminutes. The general
properties of the redshift distribution as outlined in FLY99 do not change in
any major way with the use of the new templates, but for the already
mentioned correction of the systematic problem between $z=2$ and $z=4$. As a
reference, it must be remarked that FLY99 was also a refinement of our
previous work, LYF96, written at a time when the infrared images were not yet
available.

\subsection{Cross-Identification of Objects in Both Catalogs}

  To match objects in both catalogs, we cross-correlate the coordinates and
compare apparent magnitudes, which proves to be sufficient to unequivocally
pair 98\% of the objects in the spectroscopic redshift catalog with objects
in the photometric redshift catalog. Only in three cases was it necessary to
use the redshift information, as two objects with different photometric
redshifts were being associated to a single object in the C00 list. In these
cases (see Table 1) the object whose photometric redshift was closer to the
observed spectroscopic redshift was taken to be the counterpart. It may be
claimed that this method will produce a slight bias favoring the goodness of
the comparison. It must be remarked, though, that in these cases the position
listed by C00 (and supposedly in that case, the position of the center of the
slit when the observations were made) is actually half way in between the two
objects in our catalog, which are in every case approximately equally
bright. Under these circumstances, the decision as to which object is the one
actually observed can only depend on extra information, the best available to
us being precisely the one given by the photometric redshift. Table 2 lists
the complete cross-identifications.

  The two objects listed by C00 as coming from \cite{Stern99} have very clear
counterparts in our catalog. In fact, both of them were actually passed by
ourselves to the authors as objects with $z \approx 5$ to be observed with
Keck. Their positions as listed by C00, though, differ from the positions in
our catalog by more than 1 arcsecond, and needed to be matched individually
after the original cross-identification.

  For the sake of completeness, we also include in Table 3 the objects listed
in Table 10 of C00. These are the only objects with $R<24$ for which the
authors could not determine a redshift, although they have spectra of quality
comparable to others in the sample. Their magnitudes range between $R=23.2$
and $R=24.0$. One of them (HDF36526\_1202) does not have a counterpart in our
FLY99 catalog. After careful checking we have discovered that this object
(for which we measured its photometry and redshift, see Table 4) did not
enter the final published version of the catalog, although it fulfills all
the positional and brightness criteria that we used. Another object listed in
Table 10 of C00 (HDF36378\_1235) does not lie in the area we studied, so it
is not included in our Table.\footnotemark \footnotetext{This object has---to
less than 0.5 arcseconds and 0.1 magnitudes---the same position and
brightness of the object mentioned in the previous subsection as the one
whose coordinates put it out of the HDF as defined by us, which {\it does
have a redshift} ($z=0.485$). It is one of the objects added as a ``note
added in proof'' by Cohen and collaborators, which may explain this apparent
mistake.}

  The final comparison catalog includes 146 objects with reliable
spectroscopic and photometric redshifts.

\section{COMPARISON OF PHOTOMETRIC AND SPECTROSCOPIC REDSHIFTS}

  In this section we present a direct comparison of redshift estimates based
on spectroscopic and photometric methods for all 146 objects in the combined
catalog. This is, to the best of our knowledge, the most accurate comparison
ever performed of the two techniques for this kind of deep sample, because it
includes {\it all} the spectroscopic redshifts available, as obtained by
several groups of observers using the largest telescopes in operation, over a
period of over four years, and with several values retracted and/or ammended
when known to be in error. Moreover, as the authors of C00 remark, the
redshifts given in their tables are not {\it ``preliminary values,''} but
{\it ``final redshifts.''}

  Table 2 and Figures 1 and 2 show the comparison of the redshifts assigned
by the spectroscopic and photometric methods to 146 objects in the HDF.  

  We used a $4\sigma$-clipping algorithm to measure the dispersion between
the redshift values measured with both techniques---with a sample of 146
objects, the probability of having one $>4\sigma$ discordant point by chance
is less than 1\% for a normal distribution. Our statistic of choice is
$\Delta z/(1+z)=(z_{\rm phot}-z_{\rm spec})/(1+z_{\rm spec})$. When the
sample is taken as a whole (146 objects), the mean value is $\langle\Delta
z/(1+z)\rangle=0.0035$, with rms dispersion $\sigma_{\Delta z/(1+z)}=0.065$,
and nine objects having discordant redshifts---over the four sigma
level. When we break the sample into three subsamples (``low'' redshift from
0.0 to 1.5 with 113 objects, ``medium'' redshift from 1.5 to 4.0 with 27
objects, and ``high'' redshift from 4.0 to 6.0 with 6 objects), the
respective mean values are $\langle\Delta z/(1+z)\rangle=$ 0.0010, 0.0160,
and $-0.0045$\footnotemark \footnotetext{Notice that there is an
observational ``void'' in the spectroscopic redshift distribution with no
objects with $1.355<z_{\rm{spec}}<1.980$. Objects at redshifts $1.0-1.4$
and $2.0-2.5$ are well represented in our sample, though}. The rms
dispersions are $\sigma_{\Delta z/(1+z)}=$ 0.063, 0.076, and 0.016, and the
numbers of discordant points are, respectively, 7, 2, and 0. All these values
are tabulated in Table 5. Figure 3 shows that the distribution of the
residuals, except for the $>4\sigma$-discordant points mentioned above, is
very well represented by a gaussian. It also supports the choice of a
$4\sigma$ value for the clipping algorithm.

  The first immediate result from this comparison confirms what was known
previously: careful enough photometric redshift techniques, when applied to
data of good enough quality (in terms of accurate photometry and realistic
error estimates), are able to measure redshifts to an accuracy better than
$\sigma_{\Delta z/(1+z)} \approx 0.10$, with less than 10\% of the
measurements being discordant.  In fact, {\it in a first assessment} our
particular application of these techniques reaches an accuracy better than
$\sigma_{\Delta z/(1+z)} \approx 0.08$ over the whole observed range ($z=0$
to $z=6$), with less than 7\% discordances. This result is in perfect
agreement with our previously published estimates in FLY99 and Y00.

  When assessing the errors associated to the photometric redshift
measurements, we must take into account that they are twofold: on one side,
the fact that we are using a discrete, finite, set of SEDs will produce a
systematic error, as obviously not all galaxies (in fact, {\it no galaxy})
can be exactly represented by them. We refer to this source of systematic
error as ``cosmic variance'', and it is most important in the bright regime,
when photometric errors are negligible. On the other hand, when very faint
galaxies are studied, the cosmic variance becomes negligible and the errors
in the photometric redshift induced by the uncertainties in the galaxy
photometry dominate the error budget. The photometric component of the error
can be assessed using the likelihood function, as the photometric
uncertainties are included in its calculation. The assessment of the
systematic errors, however, must be attained by the comparison of photometric
and accurate, reliable, spectroscopic redshifts for a suitably large sample
of bright galaxies---where the photometric errors will not dominate. In order
to ensure that we have such a sample, we study in the next section the
objects in our catalog that show discordances larger than $4\sigma$.

\section{THE DISCORDANT 7\%: A BLIND TEST OF SPECTROSCOPIC REDSHIFTS}

  In this section we present in detail the available data for the nine
objects that show discordant redshifts in the photometric and spectroscopic
catalogs. The photometric data include the fluxes in 7 bands (HST F300W,
F450W, F606W, and F814W; and ground-based $J$, $H$, and $K$), from which our
photometric redshift technique yields the estimated redshifts, best-fit
galaxy types, and redshift likelihood functions. Regarding the spectroscopic
data, we have looked at the original sources of publication whenever
available, and studied carefully the spectra presented. Most of the objects
scrutinized in this section, however, are new additions by Cohen and
collaborators who do not present the spectra in their paper.

  Together with a brief discussion on each object, we also present in Figure
4 the photometric data (filled points with vertical error bars indicating the
1-$\sigma$ errors and horizontal error bars showing each filter FWHM). The
figure also shows in the central panel our best-fit spectrum obtained by
taking the best-fit redshift and type, and applying the \ion{H}{1} absorption
as described in LYF96 and FLY99, the fluxes expected from our best-fit
spectrum through all seven filters (empty circles), and the redshift
likelihood function for each object. In the top panels we show the spectra of
our six galaxy templates (corresponding from left to right and top to bottom
to E/S0, Sbc, Scd, Irr, SB1 and SB2, see FLY99 and Y00) redshifted to the
spectroscopic redshift and with the effect of intergalactic \ion{H}{1}
absorption added. In most cases, that is the best approximation to the
``observed spectrum'' we can get.

  Thumbnail F814W images of the objects with discordant redshifts are
presented in Figure 5. Table 6 presents all the relevant information about
each of these objects.

  Before entering the individual discussions, we would like to remark on two
important facts regarding the expected nature of the errors in both methods.
First, the spectroscopic redshifts must be extremely precise by their own
nature---to within $\sigma_z \approx 0.001$ for low-medium resolution
spectroscopy (C00 quote $\approx 200 \kms$ as ``typical uncertainty''). In
case of error, it would be due to misidentification of absorption and/or
emission features, and the redshift values should not show any particular
regularity and/or concentration. On the other hand, photometric redshifts
have a built-in lack of precision due to the broad-band nature of the
features analysed. This leads to a rms dispersion of (in our case)
$\sigma_{\Delta z/(1+z)} \apl 0.08$. In this case, however, the confusion
between different redshifts is fully characterized by the known colors of the
redshifted templates, and the likelihood of any redshift value will always be
reflected in the likelihood function---as long as the collection of SEDs used
in the analysis really covers all possible spectral types.  These
considerations lead us to the following assertions:

\begin{description}
\item[a)] When the redshifts obtained by both techniques disagree, but the
likelihood function shows a secondary peak at (or near) the value of $z_{\rm
spec}$, then it is most likely that the spectroscopic redshift is correct.
Detailed analysis of the likelihood function must be performed in these cases
in order to obtain a ``confidence interval'' around the photometric redshift
value and, if still necessary, discover the origin of the discordance.

\item[b)] When there is no secondary peak in the likelihood function at (or
near) the value suggested by the spectroscopy, and there are spectral
features that match the redshift measured by the photometric redshift
technique, it is most likely that the photometric redshift is correct.  In
this case, the spectroscopic redshift may be wrong due to misidentification
of spectral features.

\item[c)] When both techniques disagree and there is no hint either in the
spectrum or in the likelihood function about which one is wrong, all the
evidence must be produced and the possible causes for the disagreement
analyzed. These include, but are not limited to: observational errors (like
misassignment of spectra to objects), presence of nearby objects that may
alter the accuracy of the photometry or send light into the slit during
spectroscopy, and the possibility of the object belonging to a spectral class
not included in the photometric templates (\eg a QSO or a star).
\end{description}

  With these ideas in mind, we start the individual study of all nine objects
with discordant redshifts. Where no explicit information is given about the
original source of the spectroscopic redshifts, they came from Cohen and
collaborators' new data. We follow here the naming convention that assigns
the name HDFXXXXX\_YYYY to the object with coordinates (J2000) 
$12^h XX^m XX\fs X$, $+62\degr YY\arcmin YY\arcsec$.

\subsection{HDF36386\_1234}

  This is an object of magnitude $R=24.04$. The spectroscopic redshift is
0.904, and is listed as having quality class 9, which means it has been
assigned a redshift based solely on a {\it ``single strong absorption
feature, assumed to be 2800 \AA\ because of the shape of the continuum''}
[see \cite{Cohen99} for a full description of the spectroscopic quality
classes used by the authors]. This means that a strong absorption at
approximately 5330 \AA\ must be the only clear spectroscopic feature of this
object.

  Our technique assigns to this object (number 727 in FLY99) a redshift of
0.15. It is clear from Figure 4 that the very low flux in the F300W band is
very difficult to reconcile with the shape of the rest of the spectrum if the
galaxy is at $z \approx 0.9$.  This fact reduces the likelihood of
$z\approx0.9$ to negligible values. On the other hand, to explain the 5330
\AA\ feature using our redshift we would have to identify it with the G band,
which would assign to this object a redshift $z \approx 0.24$, or with
H$\beta$ at $z \approx 0.10$. Both would be perfectly in line with our
typical dispersion, but this identification is not at all definitive.

  As can be observed in Figure 5, there is a nearby object in projection on
the sky, less than three arcseconds away. The redshift of this object is
(C00) $z_{\rm spec}=0.944$. We cannot discard the possibility of confusion
between both objects.

  Summing up, we consider this object {\it unlikely} to be at the suggested
$z_{\rm spec}=0.904$ due to its SED, but cannot present evidence conclusive
enough to definitively prove that the photometric value $z_{\rm phot}=0.15$
is right.

\subsection{HDF36396\_1230}

  This is a very interesting object. According to C00, it is a {\it
``broad-line AGN''}, given that its brightness ($R=24.4$ at $z_{\rm
spec}=0.943$) corresponds to only a few times $L^*$, too low to be a QSO. It
is also identified as the brightest object in the peak (cluster?) of the
redshift distribution at $z\approx0.96$. It is assigned spectroscopic quality
7, which indicates {\it ``only one broad emission line, assumed to be 2800
\AA''} (which means it lies at $\approx 5440$ \AA\ in the observer's
frame). In Figure 4 we show the spectrum of an average QSO template (Chen
2000, private communication) instead of all six galaxy templates at $z=z_{\rm
spec}$, to take its claimed nature into account. Cohen \etal also mention in
Section 3 of their paper that they did not include this object in their {\it
``blind check of photometric redshifts''} because {\it ``none of the groups
came close to predicting [its] redshift''}.

  As can be seen immediately, the SED of a QSO at $z=0.943$ looks {\it
absolutely different} from what is observed in this case. This fact alone
leads us to claim that the likelihood of this object being a $z=0.943$ QSO is
negligible. The likelihood function shows that the case of either a normal or
a starbursting galaxy at the same redshift can also be discarded with the
same level of security---unless the spectrum is a pathological case, and
cannot be represented by any ``normal'', ``starburst'', or QSO spectrum, the
redshift is unlikely to be $z \approx 0.94$. Figure 5 shows that our 
photometry should not be contaminated by bright nearby neighbors.

  If we study now our photometric redshift (object number 688 in FLY99), we
find a perfect agreement for the SED with one corresponding to a Scd galaxy
at redshift $z_{\rm phot}=3.40$.  At this redshift, the Lyman $\alpha$ line
would be redshifted to 5350 \AA.

  Combining the width of the peak in our likelihood function with the
detection (as indicated by C00) of a strong emission line at $\lambda
\approx$ 5440 \AA, and the presence of a strong break in the flux of the
object at approximately the same wavelength, we are led to interpret this
object as being a normal galaxy at $z \approx 3.475$, in perfect agreement
with our estimate. We cannot elaborate on the broad component of the emission
line with the data available to us.

\subsection{HDF36409\_1205}

  This is yet another object of spectral quality class 9, with spectroscopic
redshift $z_{\rm spec}=0.882$ and magnitude $R=22.94$. It is listed by C00 as
having appeared for the first time in C96, but we have not been able to
locate it in that reference. The absorption feature putatively identified by
C00 as Mg{\rm II} 2800 must be at $\lambda=5270$ \AA. Figure 4 shows a case
very similar to that of HDF36386\_1234: there is an obvious lack of flux in
the F300W filter, that renders unlikely---in our analysis---the $z\approx0.9$
interpretation.

  With our photometric redshift technique, we obtain an excellent fit to the
SED of this object for a $z_{\rm phot}=0.00$ Scd galaxy. The only tentative
interpretation of the observed spectrum that is consistent with this is the
observed absorption feature corresponding to Mg {\rm I} absorption at
$z=0.02$.

  Once again in this case, we are forced to conclude that the spectroscopic
suggested value $z_{\rm spec}=0.882$ is {\it unlikely} to be the true
redshift, but we have no other strong evidence to support our case for
$z_{\rm phot}=0.00$ either. Moreover, it can be seen in Figure 5 that this
object lies in a particularly crowded region of the sky, which makes the case
all the more difficult.

\subsection{HDF36414\_1143}

  This object is listed in C00 as having magnitude $R=23.52$ and redshift
$z_{\rm spec}=0.548$. Quality class is 3, meaning {\it ``multiple features,
faint, id uncertain''.}

  Looking at Figure 4 we see that the SED does not fit any type of galaxy at
this redshift, and that our interpretation of it as a Scd galaxy at redshift
$z_{\rm phot}=1.32$ is, while better, not optimal. One possible reason for
this is that this object (number 200 in FLY99, $AB(8140)=23.19$) actually
overlaps in the sky with another {\it brighter} object (number 183 in FLY99,
$AB(8140)=22.42$) with a spectroscopic redshift listed in C00 as $z_{\rm
spec}=0.585$. The distance between the centers of both objects is only 2
arcseconds.

  If confusion caused by the proximity of the second object is not the issue
here, then we should try to identify some of the features described by C00
with absorption features at approximately $z=1.32$. If, for example, Cohen
and collaborators had identified an absorption feature as Ca {\rm II} (H,K)
at $z=0.548$ (which would put it at $\lambda \approx$ 6120 \AA), it could
also be Mg {\rm II} at $z \approx 1.19$. {\it Mutatis mutandis}, the G band
at $z=0.548$ could also be identified as Mg {\rm II} at $z \approx 1.38$.

  As a resume, once again we have to admit that we have no clear explanation
as to which is the reason for the discordance in this case, if not the
possibility of confusion induced by the vicinity of another bright object.

\subsection{HDF36441\_1410}

  This object was first presented in \cite{L97}(L97 hereafter).  Its $R$
magnitude is 24.26, and the listed redshift is $z_{\rm spec}=2.267$.  C00
assesses it as a ``secure'' redshift, while Lowenthal and collaborators
classify it as ``definite.'' The authors of L97 kindly published their
spectra in Figure 2 of their paper, so we can assess by ourselves the quality
of the spectrum.

  Our analysis assigns to this object a redshift of 0.01. It can be seen in
Figure 4 that the fit to the SED at the spectroscopic redshift is very good
using a SB1 type, in fact (to the eye) comparable to the fit to the SED
redshifted to our photometric value. The difference, though, is given by the
F606W and F814W bands having tiny error bars, so the weight of the likelihood
function is driven by them and their better fit to the photometric
value. This can of course be appreciated from the fact that there is no
secondary peak in the likelihood function near the spectroscopic value.

  After a first look at Figures 4 and 5 nothing seems to be wrong with the
spectroscopic value, and we actually counted this object as one of our
failures in FLY99. Nevertheless, encouraged by the general success of our
technique and piqued by the absence of a secondary peak at $z \approx z_{\rm
spec}$, we tried to find the expected features of a $z \approx 0.01$ galaxy
in the observed spectrum as presented in L97.

  We then found the following: the emission line identified in L97 as Lyman
$\alpha$ can actually be identified with [\ion{O}{2}] 3727 at $z=0.065$. In
that case, the Calcium doublet should be observed at $\lambda$=4190,4228 \AA,
the G absorption band should be at $\lambda=$4580 \AA, and the Na {\rm I}
absorption feature should be observed at $\lambda=$6278 \AA. All these
features can be observed in the spectrum with {\it at least} the same level
of significance of most of the original identifications. Lowenthal and
collaborators considered this identification as a possibility in their paper
(see Appendix in L97), and decided from the implied luminosity and equivalent
width of the putative [\ion{O}{2}] line that it was ``not implausible.''

  It should also be noticed that the presence of a continuum break expected
at the position of the Lyman $\alpha$ line as identified in L97 (which would
be identified as the 4000 \AA\ break according to the photometric redshift)
cannot be checked in the spectrum, because it would lie at the very blue end
of it.  Also, the slope of the UV spectrum seems too steep for a typical
$z\approx2$ galaxy, as can be seen by comparing it with other spectra in the
same figure in L97.

  Our conclusion about this object is that the photometric value fits better
all the available evidence, including the spectrum itself. The spectroscopic
value is likely to be in error due to misidentification of features in a
noisy spectrum.

\subsection{HDF36478\_1256}

  This is another object described originally in L97. It has $z_{\rm
spec}=2.931$ and $R=24.35$.  The spectroscopic quality is 2 ({\it ``good with
multiple features''}) in C00, and ``definite'' in the original reference. As
can be seen in the top panels of Figure 4, its SED can be almost perfectly
fitted by an Irr galaxy at the spectroscopic redshift, with the only slight
caveat being the presence of detectable but faint ($\approx 3\sigma$) flux in
the F300W filter.

  In a very similar fashion to the previous case, our photometric method
prefers a lower redshift solution at $z_{\rm phot}=0.26$. As above, the break
in the spectrum which corresponded to the onset of the Lyman $\alpha$ forest
using the redshift value from L97 is then taken to be the 4000 \AA\ break.

  However, detailed observation of our likelihood function shows that there
is a second peak at $z \approx 3$. The secondary peak actually reaches a
maximum at $z=3.13$, with a value which is almost 20\% of the maximum at
$z=0.26$.

  Following our previously stated considerations, we deem our technique gives
us in this case an incorrect answer, as the spectroscopic evidence agrees
within the errors with a secondary peak in the likelihood function.
Nevertheless, when the likelihood function is used to perform a complete
assessment of the errors in $z_{\rm phot}$, it shows the value $z=2.931$ to
be ``within the errors'': the $1\sigma$ interval around $z_{\rm phot}$ is
0.15--0.38, the $2\sigma$ interval is 0.06--0.52 {\it plus} 2.80--3.49.

\subsection{HDF36494\_1317}

  This object is listed in C00 as having $z_{\rm spec}=0.271$, with a
magnitude $R=23.63$ and a quality assessment of 3.

  Figure 4 shows that, while there is no way that a redshift 0.271 SED can
fit the observed photometry, a perfect fit is achieved by using a Scd
spectral template at $z_{\rm phot}=1.24$.

  A footnote in C00 refers to this object with the following text: {\it
``Definite emission line at 8340 \AA, possible emission line at 6363 \AA. If
both are real, $z=0.271$. If only the stronger one is real, then
$z=1.238$. Spectrum too red to reach 3727 \AA\ if $z=0.271$.''}

  With this information in hand, we consider we can peacefully rest our
case. The redshift of the galaxy is perfectly well measured to be $z=1.238$,
in exact agreement with our technique.

\subsection{HDF36561\_1330}

  An object with $z_{\rm spec}=0.271$, $R=23.80$, and spectral quality
4. Figure 4 shows that a fit can be obtained at that redshift for the
photometry, though the goodness-of-fit is not comparable with the one 
obtained using the photometric redshift technique.

  The photometric redshift is $z_{\rm phot}=1.07$, with the likelihood 
function excluding any other possibility for the redshift. 

  In this case, as in the previous one, a footnote in C00 adds more
information: {\it ``Emission line at 8340 \AA\ is interpreted as
H$\alpha$. Spectrum too red to reach 3727 \AA\ if $z=0.271$.'' }

  As in the previous case, we think that the observed line must be identified
as being O {\rm II} 3727 at a redshift $z=1.238$, which will be in agreement
with our photometric measurement to $\approx $ 8\% error in $\Delta
z/(1+z)$.

\subsection{HDF36569\_1302}

  This object is listed in C00 as having magnitude $R=23.84$, redshift
$z_{\rm spec}=0.474$, and spectral quality 1 ({\it ``secure redshift,
multiple features identified''}). In Figure 4, though, we see that the
spectrum is much too bright in the IR bands, and that none of the SEDs can
actually fit the observations when redshifted to that value. Examination of
Figure 5 allows us to discard the possibility of any nearby IR-bright object
interfering with our measurements.

  Our photometric redshift technique suggests a value of 1.27, with a
spectral type Sbc. No secondary peaks are apparent in the likelihood
function. However, we have tried to cross-identify the possible features that
the observers may have identified to be at $z=0.474$, and found no way to
position them on a $z\approx1.27$ spectrum.

  Careful observation of the circumstances is thus required. A look at Figure
5 shows that this object is close ($<3.5\arcsec$) to a very bright
($R=21.07$) galaxy, number 458 in FLY99, already observed with Keck by C96
and the Hawaii group. This bright galaxy has a spectroscopic redshift of
0.475, as listed in C00, and has an SED which is utterly different from the
one of HDF36569\_1302. No sign of interaction or merging is visible, which
could be used to explain the same redshift being assigned to both objects.

  We are led to conclude in this case that the spectroscopic redshift is in 
error due to confusion with a nearby bright source. Given that the only 
evidence left is the photometry, and that our technique has shown to be right 
in over 95\% of the cases, we conclude that the redshift of this object must
be (within errors) $z=1.27$.

\section{AN EXTRA POSSIBLY WRONG SPECTROSCOPIC REDSHIFT}

  We present in this Section the photometric and spectroscopic data for an
extra object. Although in this case the redshift difference is not over
$4\sigma$ (our stated criterium to define ``discordance''), we have good
reasons to believe the spectroscopic value may be in error.

  The object in question is HDF36450\_1251. It is completely isolated (see
Figure 5), and C00 lists for it a redshift $z_{\rm spec}=2.801$ and a
magnitude $R=24.37$. The authors of C00 assign to the spectrum a quality flag
3, which means there are multiple identified features, but their
identification is uncertain.  

  One problem springs immediately into view when looking at Figure 4---the
object has an $11\sigma$ detection in the F300W image, where zero flux would
be expected from an object at such a high redshift.  Our photometric
technique suggests a lower redshift $z_{\rm phot}=2.23$. The SED fits
perfectly, and the likelihood function admits quite a wide range of values,
essentially between $z=2.1$ and $z=2.3$. The F300W detection makes the
likelihood of the object being at any $z>2.6$ completely negligible.

  We conclude in this case that the redshift assigned by the spectroscopic 
method cannot be right due to serious discordance with the observations.

\section{DISCUSSION}

  We have presented in this paper a detailed comparison of the cleanest,
deepest and most complete catalog of spectroscopic redshifts available for
the HDF with our catalog of photometric redshifts. The first result of the
comparison is that the photometric redshifts are accurate to within $\Delta
z/(1+z)=0.08$ (rms) {\it at all redshifts explored between $z=0$ and
$z=6$}. The fraction of {\it apparently} discordant points is 6.2\%. This
fraction, when taken in separate redshift intervals, amounts to 6.2\% at
$z=0.0-1.5$ (7 objects out of 113), 7.4\% at $z=1.5-4.0$ (2 objects out of
27), and 0\% at $z>4.0$ (out of six objects).

  We have performed a careful study of all the available evidence regarding
the discordant redshifts. This analysis has resulted in the following:

\begin{itemize}
\item In five out of nine cases, we have evidence that indicates that {\it
the spectroscopic redshifts are in error.} The causes for this range from
misidentification of spectral features in noisy spectra to confusion with
nearby objects.

\item Only in one of the nine discordant cases we have concluded that the
photometric value is not correct. A full study of the errors associated with 
the value $z_{\rm phot}$, though, shows that the spectroscopic and photometric 
values {\it do agree} to within a $2\sigma$ confidence level.

\item In three cases the evidence we have gathered is not enough to decide
which of the techniques is giving more accurate results. We kindly ask from
these lines the authors of C00 to produce the evidence, to help in deciding
on these cases.
\end{itemize}

  Once this analysis is fully taken into account, the final figures for the
accuracy of our redshift determination technique become the following: The
rms dispersions in $\Delta z/(1+z)$ at low, medium, and high redshift are
$\sigma_{\Delta z/(1+z)}=$ 0.063, 0.068, and 0.016, with mean values
$\langle\Delta z/(1+z) \rangle=$ 0.0003, 0.0186, and $-0.0045$, and
discordant fractions 0\%, 0\% ({\it outside the quoted error bars}), and
0\%. The total number of points in each redshift bin are 109, 26, and 6,
respectively.  All these final values are also tabulated in Table 5, and
presented in Figures 6 and 7.

  This result shows that {\it at most} three out of 146 galaxies can have
SEDs which are not well represented in our template set. Even in the case
that in all three cases the photometric redshifts were shown to be in error,
and that these errors were exclusively caused by the incompleteness of our
SED sample (leaving aside other possible systematic effects induced by
variance of HI absorption along different sightlines or other problems that
may be inherent to our likelihood technique), this would set a stringent
limit to the percentage of galaxies which are not well represented by our
templates. Applying Poisson statistics to these figures, we get that such a
percentage cannot be (to a $2\sigma$ confidence level) larger than 5.4\%, and
will be smaller (and possibly zero) if any or all of those three uncertain
cases are shown to be caused by errors in the spectroscopic
measurements. This result holds also when the putative presence of extra dust
in the observed spectra is taken into account. We detect no evidence in our
analysis for the presence of highly-reddened galaxies whose SEDs cannot be
reproduced with our templates (which, being observational in origin, {\it do}
contain some amount of dust).

  We can use this sample of photometric and spectroscopic redshifts to
characterize the errors associated with our technique. A full description of
this procedure will be presented elsewhere, but we describe here the
operational steps necessary to measure the confidence intervals for any
particular object. As was stated in the Introduction, and also in
\cite{Lanzetta98} and Y00, the photometric redshift error has two components.
The first one is induced by photometric uncertainty in the measurement of the
fluxes, dominates at the faint end, and is fully taken into account by the
use of the likelihood function. The second one is systematic, due to cosmic
variance, and can be characterized by a typical dispersion measured from a
sample of bright objects, where the photometric error is negligible. Using
the sample we have just presented, we estimate (as a zero-th order
approximation) this dispersion to be constant in $\Delta z/(1+z)$, and equal
to $\sigma_{\Delta z/(1+z)} = 0.065$.

  In order to combine both error components, we can convolve the redshift
likelihood function with a gaussian with a variable $\sigma_G(z) = (1+z) \times
\sigma_{\Delta z/(1+z)}$. In doing this we are assuming that the systematic
error follows a normal distribution, which was tested in Section 3. The
resulting function can be normalized to yield a redshift probability density,
which can be used following the standard methods to calculate confidence
intervals. In those cases where the probability density is multimodal---as is
the case with HDF36478\_1256---the confidence intervals may also be
multimodal, \ie disjoint. This procedure, when applied to the particular case
of HDF36478\_1256, yields the above stated confidence limits (0.15--0.38,
$1\sigma$) (0.06--0.52 {\it plus} 2.80--3.49, $2\sigma$). In most bright 
objects, though, the redshift likelihood function shows a single, sharp 
peak. In these cases the convolution with the gaussian will obviously 
produce a curve that is very approximately gaussian, and the confidence 
limits will approximately coincide with those given by the value of 
$\sigma_G(z)$ used.

  It is a very remarkable fact that our technique has an ``error rate'' of
0\% (in number of confirmed wrong redshifts compared to total), which cannot
be higher than 2\% even if all three dubious cases prove to be wrong. In
comparison, the spectroscopic technique error rate is {\it at least} 3.4\%
(even when only {\it ``definitive''} redshifts, as published by C00, are
taken into account, and assuming all three undefined discordant redshifts
have correct values of $z_{\rm spec}$), and it can be {as high as} 14.3\% or
higher (when all the HDF spectroscopic values that have been published and
then retracted are included, together with those objects which have been
observed but for which a spectroscopic redshift has been impossible to
measure).

\section{CONCLUSIONS}

  We have shown that our photometric redshift technique produces accurate
measurements to $\Delta z/(1+z) \apl 0.07$ (rms) at all redshifts $0<z<6$. We
have also shown that in every case where there is a large difference between
carefully measured spectroscopic redshifts and carefully measured photometric
redshifts, and the systematic and photometric errors are included in the
analysis, the spectroscopic redshift is the one in error (although in a few
cases we have no evidence to prove either one to be right). In fact, the
proportion of incorrect spectroscopic redshifts {\it even after careful
double-checking of all the data involved} is $\approx$ 3\% and, depending on
how they are counted, it may be as high as 14\%. On the other hand, in the
single studied case where the photometric value seems to be incorrect, the
analysis of the likelihood function shows that the real redshift value is
indeed within the errors given by the photometric method.

  One conclusion we extract from this result is that our set of spectral
energy distribution templates is a fair, complete, dense representation of
all the galaxy spectral distributions that have been observed to date. We can
confirm that in the bright limit, where the photometric errors can be
ignored---this applies in the case of the HDF to all galaxies for which
spectroscopy is available---the errors in $\Delta z /(1+z)$ are dominated by
the systematics and are {\it at all redshifts} $\apl 0.07$ (rms). This paper
further proves that there is a strict upper limit (less than 5.4\%, to
$2\sigma$ confidence) to the possible percentage of galaxies whose SEDs
differ significantly from our templates in a flux limited ($R<24$) sample.
This limit is calculated by applying Poisson statistics to the fact that {\it
at most three galaxies out of 146} may have redshifts which disagree with our
measurements to a significant (more than $4\sigma$) level.

  In the limit of bright objects the likelihood function (which by its nature
only includes the photometric information) has to be convolved with an
(assumed) gaussian distribution with the above calculated $\sigma_{\Delta
z/(1+z)}$ in order to account for the systematic errors and to obtain
realistic errors. In the faint limit, where the photometric errors dominate,
the likelihood function by itself represents a good assessment of the
errors---although, of course, it must also be convolved with the same
gaussian in order to include the systematic component.

  There is an important conclusion that can be extracted from this work that
affects other photometric redshift techniques. Some of the methods that have
been put forward [\eg \cite{Wang98}; \cite{Connolly95}; \cite{Csabai00}] are
based on the use of a ``calibrating sample'' of redshifts, which are taken to
be perfectly measured. Using this sample a set of parameters is determined
that gives the most accurate possible reproduction of the observed
spectroscopic redshifts, and then these parameters are used to calculate
redshifts of objects which do not have spectroscopic values. One of the most
popular versions of this method uses polynomial fits to define a ``redshift
vs observed colors'' function [like in \cite{Wang98} and \cite{Connolly95}].
Another possibility is to iteratively ``tune'' or ``improve'' the galactic
spectral templates to minimize the dispersion in the $z_{\rm phot}$ versus
$z_{\rm spec}$ plane [this is the method followed by \cite{Csabai00}]. {\it
Both these techniques, as well as any other using similar methods, are
extremely sensitive to errors in the spectroscopic sample}. By their own
nature they will try to ``absorb'' any outlier point into the fit, hence
improving the ``formal'' fit at the price of distorting the real underlying
relation, and predicting wrong redshifts for any point with similar colors to
the ones in error.

  It is obvious that the spectroscopic redshift technique has limits in
apparent magnitude and/or signal-to-noise ratio where problems arise. Our
analysis shows that by magnitude $AB(8140) \approx 24$ even the best
instruments available produce spectra that are susceptible to line
misidentification, even when carefully analyzed by expert observers. Our
photometric redshift technique, on the other hand, has proved its accuracy
and reliability for all redshifts $0<z<6$ and magnitudes $R<24$ that have
been observed. This allows us to obtain an accurate and detailed measurement
of the systematic errors associated to it. Together with the likelihood
function---which measures for each individual galaxy the errors induced by
the photometric uncertainties---and assuming that our set of
spectrophotometric templates is complete (which has been verified for all
cases observed to date), our photometric redshift technique has the advantage
over traditional spectroscopy of giving {\it complete information} on the
measured redshift {\it and} its associated error.

  We hence conclude that our photometric method is both accurate (within its
stated errors) and reliable, and that it is in fact more accurate and
reliable than the spectroscopic method when analyzing faint galaxy data.

\acknowledgments We thank Judy Cohen and her collaborators, as well as all
the other groups involved in spectroscopy of HDF objects, for making their
results public. We also thank Bob Williams and his team at STScI for
providing us with the HDF images, Mark Dickinson for making the IR Kitt Peak
images available, and an anonymous referee for useful suggestions that have
improved the paper. A.F.S. acknowledges support from a European Union Marie
Curie Fellowship. K.M.L., S.M.P., and N.Y. were supported by NASA grant
NAGW-4422 and NSF grant AST-9624216. During the preparation of this paper we
have made extensive use of NASA's ADS Abstract Service, the LANL archive, and
the NASA/IPAC Extragalactic Database (NED) which is operated by the Jet
Propulsion Laboratory, California Institute of Technology, under contract
with NASA.

\clearpage
%
%
\begin{deluxetable}{c c c c c r c c c c c}
\tablecolumns{11}
\tabletypesize{\scriptsize}
\tablewidth{0pc}
\tablecaption{Objects with double cross identifications in the catalogs\tablenotemark{a}}
\tablehead{
\multicolumn{4}{c}{Cohen \etal 2000} & \colhead{} & \multicolumn{5}{c}{Fern\'andez-Soto \etal 1999} &
\colhead{$\theta$\tablenotemark{b}} \\
\cline{1-4} \cline{6-10} \\
\colhead{RA($-12$h)}         & \colhead{Dec($-62^{\circ}$)} & 
\colhead{$z_{\rm spec}$}     & \colhead{$R$}                & \colhead{} &
\colhead{ID}                 & \colhead{RA($-12$h)}         & 
\colhead{Dec($-62^{\circ}$)} & \colhead{$z_{\rm phot}$}     & 
\colhead{$AB(8140)$}         & \colhead{(arcsec)}}
\startdata
 36:49.95 & 12:25.9 & 1.205 & 23.71 &&  353 & 36:49.95 & 12:25.4 & 1.07 & 24.32 & 0.5 \\
          &         &       &       &&  363 & 36:49.99 & 12:26.3 & 0.01 & 24.43 & 0.5 \\
 36:52.86 & 14:08.2 & 3.367 & 24.49 &&  957 & 36:52.99 & 14:08.5 & 3.48 & 26.85 & 1.0 \\
          &         &       &       &&  959 & 36:52.91 & 14:08.5 & 0.04 & 26.32 & 0.5 \\
 36:57.51 & 12:12.1 & 0.561 & 22.62 &&    7 & 36:57.60 & 12:12.4 & 0.71 & 23.16 & 0.7 \\
          &         &       &       &&    6 & 36:57.45 & 12:11.9 & 0.91 & 23.28 & 0.5 \\
\enddata
\tablenotetext{a}{The first object listed in each pair was taken as the 
identification}
\tablenotetext{b}{Distance between the position in C00 and FLY99}
\end{deluxetable}

\clearpage
%
%
\begin{deluxetable}{c c c c c r c c c c l}
\tablecolumns{11}
\tabletypesize{\scriptsize}
\tablewidth{0pt}
\tablecaption{Comparison of both catalogs}
\tablehead{
\multicolumn{4}{c}{Cohen \etal 2000} & \colhead{} & \multicolumn{5}{c}{Fern\'andez-Soto \etal 1999} &
\colhead{} \\
\cline{1-4} \cline{6-10} \\
\colhead{RA($-12$h)}         & \colhead{Dec($-62^{\circ}$)} & 
\colhead{$z_{\rm spec}$}     & \colhead{$R$}                & 
\colhead{}                   & \colhead{ID}                 & 
\colhead{RA($-12$h)}         & \colhead{Dec($-62^{\circ}$)} & 
\colhead{$z_{\rm phot}$}     & \colhead{$AB(8140)$}         & 
\colhead{$\Delta z /(1+z)$\tablenotemark{a}} }
\startdata
 36:38.43 & 12:31.2 & 0.944 & 22.87 &&  716 & 36:38.40 & 12:31.32 & 0.70 & 22.61 &\phn\phn    $-0.13  $ \\
 36:38.61 & 12:33.8 & 0.904 & 24.04 &&  727 & 36:38.60 & 12:33.86 & 0.15 & 23.93 &\phn\phn    $-0.40^*$ \\
 36:38.99 & 12:19.7 & 0.609 & 22.14 &&  617 & 36:38.96 & 12:19.77 & 0.54 & 22.21 &\phn\phn    $-0.04  $ \\
 36:39.60 & 12:30.2 & 0.943 & 24.40 &&  688 & 36:39.56 & 12:30.49 & 3.40 & 25.40 &\phn\phn\phs$ 1.26^*$ \\
 36:40.02 & 12:07.3 & 1.015 & 22.75 &&  466 & 36:40.00 & 12:07.38 & 0.93 & 22.28 &\phn\phn    $-0.04  $ \\
 36:40.85 & 12:03.1 & 1.010 & 23.49 &&  402 & 36:40.81 & 12:03.05 & 0.98 & 23.36 &\phn\phn    $-0.01  $ \\
 36:40.94 & 12:05.3 & 0.882 & 22.94 &&  424 & 36:40.95 & 12:05.36 & 0.00 & 23.28 &\phn\phn    $-0.47^*$ \\
 36:41.24 & 12:02.9 & 3.220 & 23.94 &&  390 & 36:41.23 & 12:02.93 & 3.51 & 24.03 &\phn\phn\phs$ 0.07  $ \\
 36:41.34 & 11:40.8 & 0.585 & 21.91 &&  183 & 36:41.33 & 11:41.01 & 0.75 & 22.42 &\phn\phn\phs$ 0.10  $ \\
 36:41.43 & 11:42.5 & 0.548 & 23.51 &&  200 & 36:41.39 & 11:43.01 & 1.32 & 23.19 &\phn\phn\phs$ 0.50^*$ \\
 36:41.62 & 11:31.7 & 0.089 & 19.36 &&   85 & 36:41.61 & 11:31.82 & 0.10 & 19.76 &\phn\phn\phs$ 0.01  $ \\
 36:41.62 & 12:00.5 & 0.483 & 25.03 &&  365 & 36:41.59 & 12:00.57 & 0.23 & 24.88 &\phn\phn    $-0.17  $ \\
 36:41.70 & 12:38.7 & 2.591 & 24.32 &&  702 & 36:41.71 & 12:38.82 & 2.40 & 24.60 &\phn\phn    $-0.05  $ \\
 36:41.95 & 12:05.4 & 0.432 & 20.82 &&  391 & 36:41.93 & 12:05.43 & 0.50 & 21.01 &\phn\phn\phs$ 0.05  $ \\
 36:42.04 & 13:21.2 & 0.846 & 23.95 &&  902 & 36:42.02 & 13:21.47 & 0.93 & 23.59 &\phn\phn\phs$ 0.05  $ \\
 36:42.71 & 13:06.7 & 0.485 & 22.02 &&  817 & 36:42.72 & 13:07.13 & 0.68 & 22.19 &\phn\phn\phs$ 0.13  $ \\
 36:42.93 & 12:16.4 & 0.454 & 20.51 &&  467 & 36:42.91 & 12:16.37 & 0.45 & 20.73 &\phn\phn    $-0.00  $ \\
 36:43.16 & 12:42.2 & 0.849 & 22.34 &&  694 & 36:43.15 & 12:42.23 & 0.72 & 21.74 &\phn\phn    $-0.07  $ \\
 36:43.21 & 11:48.1 & 1.010 & 23.10 &&  195 & 36:43.18 & 11:47.99 & 1.11 & 22.45 &\phn\phn\phs$ 0.05  $ \\
 36:43.42 & 11:51.4 & 1.242 & 23.00 &&  228 & 36:43.40 & 11:51.57 & 0.98 & 23.05 &\phn\phn    $-0.12  $ \\
 36:43.61 & 12:18.1 & 0.752 & 22.56 &&  465 & 36:43.62 & 12:18.28 & 0.64 & 22.71 &\phn\phn    $-0.06  $ \\
 36:43.81 & 11:42.9 & 0.765 & 21.26 &&  122 & 36:43.79 & 11:42.88 & 0.64 & 20.87 &\phn\phn    $-0.07  $ \\
 36:43.92 & 12:40.5 & 4.540 & 26.00 &&  674 & 36:43.84 & 12:41.54 & 4.45 & 25.00 &\phn\phn    $-0.02  $ \\
 36:43.97 & 12:50.1 & 0.557 & 20.84 &&  720 & 36:43.96 & 12:50.07 & 0.54 & 20.97 &\phn\phn    $-0.01  $ \\
 36:44.07 & 14:09.8 & 2.267 & 24.26 && 1062 & 36:44.07 & 14:10.05 & 0.01 & 24.63 &\phn\phn    $-0.69^*$ \\
 36:44.10 & 13:10.7 & 2.929 & 23.84 &&  815 & 36:44.09 & 13:10.75 & 3.18 & 24.03 &\phn\phn\phs$ 0.06  $ \\
 36:44.19 & 12:40.3 & 0.875 & 23.39 &&  650 & 36:44.18 & 12:40.35 & 0.92 & 23.33 &\phn\phn\phs$ 0.02  $ \\
 36:44.20 & 12:47.8 & 0.555 & 21.40 &&  709 & 36:44.18 & 12:47.83 & 0.54 & 21.62 &\phn\phn    $-0.01  $ \\
 36:44.38 & 11:33.2 & 1.050 & 21.96 &&   17 & 36:44.37 & 11:33.15 & 0.96 & 21.26 &\phn\phn    $-0.04  $ \\
 36:44.49 & 11:42.3 & 1.020 & 24.30 &&   99 & 36:44.45 & 11:42.17 & 0.96 & 22.71 &\phn\phn    $-0.03  $ \\
 36:44.61 & 13:04.6 & 0.485 & 21.14 &&  774 & 36:44.58 & 13:04.62 & 0.67 & 21.22 &\phn\phn\phs$ 0.12  $ \\
 36:44.64 & 12:27.4 & 2.500 & 23.66 &&  517 & 36:44.63 & 12:27.42 & 2.47 & 23.73 &\phn\phn    $-0.01  $ \\
 36:44.67 & 11:49.2 & 4.580 & 26.00 &&  173 & 36:44.65 & 11:50.45 & 4.53 & 25.04 &\phn\phn    $-0.01  $ \\
 36:44.83 & 12:00.1 & 0.457 & 22.85 &&  270 & 36:44.82 & 12:00.22 & 0.36 & 22.99 &\phn\phn    $-0.07  $ \\
 36:45.01 & 12:39.6 & 1.225 & 24.53 &&  625 & 36:44.99 & 12:39.63 & 1.18 & 23.89 &\phn\phn    $-0.02  $ \\
 36:45.03 & 12:51.0 & 2.801 & 24.37 &&  712 & 36:45.02 & 12:51.02 & 2.23 & 24.08 &\phn\phn    $-0.15^*$ \\
 36:45.36 & 11:53.5 & 2.799 & 22.53 &&  175 & 36:45.34 & 11:52.74 & 3.20 & 23.24 &\phn\phn\phs$ 0.11  $ \\
 36:45.36 & 13:46.8 & 3.160 & 25.15 &&  964 & 36:45.35 & 13:47.00 & 3.19 & 25.09 &\phn\phn\phs$ 0.01  $ \\
 36:45.41 & 13:25.8 & 0.441 & 22.33 &&  876 & 36:45.40 & 13:25.97 & 0.46 & 22.61 &\phn\phn\phs$ 0.01  $ \\
 36:45.86 & 13:25.7 & 0.321 & 20.71 &&  869 & 36:45.85 & 13:25.87 & 0.41 & 20.93 &\phn\phn\phs$ 0.07  $ \\
 36:45.86 & 14:11.9 & 2.427 & 24.85 && 1054 & 36:45.88 & 14:12.10 & 2.41 & 25.21 &\phn\phn    $-0.00  $ \\
 36:45.88 & 11:58.3 & 5.600 & 27.00 &&  213 & 36:45.89 & 11:58.26 & 5.65 & 27.17 &\phn\phn\phs$ 0.01  $ \\
 36:45.96 & 12:01.3 & 0.679 & 23.88 &&  247 & 36:45.95 & 12:01.40 & 0.76 & 23.72 &\phn\phn\phs$ 0.05  $ \\
 36:46.13 & 12:46.5 & 0.900 & 22.86 &&  653 & 36:46.12 & 12:46.50 & 0.59 & 22.36 &\phn\phn    $-0.16  $ \\
 36:46.17 & 11:42.2 & 1.013 & 21.52 &&   45 & 36:46.16 & 11:42.09 & 1.04 & 21.26 &\phn\phn\phs$ 0.01  $ \\
 36:46.34 & 14:04.6 & 0.962 & 21.69 && 1018 & 36:46.35 & 14:04.68 & 0.85 & 21.24 &\phn\phn    $-0.06  $ \\
 36:46.50 & 14:07.5 & 0.130 & 23.87 && 1029 & 36:46.53 & 14:07.60 & 0.16 & 23.95 &\phn\phn\phs$ 0.03  $ \\
 36:46.51 & 11:51.3 & 0.503 & 22.00 &&  124 & 36:46.50 & 11:51.33 & 0.41 & 21.99 &\phn\phn    $-0.06  $ \\
 36:46.51 & 12:03.5 & 0.454 & 24.32 &&  245 & 36:46.54 & 12:03.12 & 0.50 & 24.35 &\phn\phn\phs$ 0.03  $ \\
 36:46.80 & 11:44.9 & 1.060 & 24.23 &&   50 & 36:46.86 & 11:44.84 & 1.06 & 23.23 &\phn\phn\phs$ 0.00  $ \\
 36:46.95 & 12:26.1 & 2.969 & 25.13 &&  444 & 36:46.92 & 12:26.07 & 2.69 & 25.24 &\phn\phn    $-0.07  $ \\
 36:47.00 & 12:36.9 & 0.321 & 20.62 &&  537 & 36:47.03 & 12:36.86 & 0.47 & 20.98 &\phn\phn\phs$ 0.11  $ \\
 36:47.10 & 12:12.5 & 0.677 & 24.63 &&  318 & 36:47.07 & 12:12.52 & 0.75 & 24.82 &\phn\phn\phs$ 0.04  $ \\
 36:47.16 & 14:14.4 & 0.609 & 23.92 && 1048 & 36:47.18 & 14:14.24 & 0.69 & 23.55 &\phn\phn\phs$ 0.05  $ \\
 36:47.17 & 13:41.7 & 1.313 & 23.93 &&  917 & 36:47.17 & 13:41.88 & 1.16 & 23.72 &\phn\phn    $-0.07  $ \\
 36:47.28 & 12:30.7 & 0.421 & 22.63 &&  476 & 36:47.28 & 12:30.68 & 0.47 & 22.77 &\phn\phn\phs$ 0.03  $ \\
 36:47.55 & 12:52.7 & 0.681 & 24.26 &&  671 & 36:47.54 & 12:52.68 & 0.75 & 23.93 &\phn\phn\phs$ 0.04  $ \\
 36:47.75 & 12:55.7 & 2.931 & 24.35 &&  687 & 36:47.77 & 12:55.68 & 0.26 & 23.95 &\phn\phn    $-0.68^*$ \\
 36:47.79 & 12:32.9 & 0.960 & 23.80 &&  483 & 36:47.78 & 12:32.94 & 1.03 & 23.34 &\phn\phn\phs$ 0.04  $ \\
 36:48.07 & 13:09.0 & 0.476 & 20.43 &&  751 & 36:48.07 & 13:09.02 & 0.38 & 20.45 &\phn\phn    $-0.07  $ \\
 36:48.27 & 14:17.1 & 2.005 & 23.58 && 1044 & 36:48.32 & 14:16.61 & 2.49 & 23.40 &\phn\phn\phs$ 0.16  $ \\
 36:48.29 & 14:26.3 & 0.139 & 18.70 && 1067 & 36:48.32 & 14:26.25 & 0.11 & 19.18 &\phn\phn    $-0.03  $ \\
 36:48.33 & 11:46.0 & 2.980 & 24.58 &&   21 & 36:48.29 & 11:45.91 & 3.17 & 25.07 &\phn\phn\phs$ 0.05  $ \\
 36:48.33 & 12:14.3 & 0.962 & 23.87 &&  297 & 36:48.25 & 12:13.91 & 1.12 & 22.55 &\phn\phn\phs$ 0.08  $ \\
 36:48.62 & 13:28.1 & 0.958 & 23.14 &&  838 & 36:48.62 & 13:28.23 & 1.06 & 22.74 &\phn\phn\phs$ 0.05  $ \\
 36:48.77 & 13:18.4 & 0.753 & 22.87 &&  780 & 36:48.78 & 13:18.45 & 0.79 & 22.69 &\phn\phn\phs$ 0.02  $ \\
 36:48.98 & 12:45.8 & 0.512 & 23.48 &&  565 & 36:48.99 & 12:45.89 & 0.53 & 23.66 &\phn\phn\phs$ 0.01  $ \\
 36:49.05 & 12:21.1 & 0.953 & 22.59 &&  339 & 36:49.05 & 12:21.24 & 0.85 & 22.44 &\phn\phn    $-0.05  $ \\
 36:49.24 & 11:48.8 & 0.961 & 23.26 &&   18 & 36:49.24 & 11:48.49 & 0.99 & 22.95 &\phn\phn\phs$ 0.01  $ \\
 36:49.34 & 11:55.1 & 0.961 & 23.36 &&   79 & 36:49.33 & 11:55.05 & 0.95 & 23.43 &\phn\phn    $-0.01  $ \\
 36:49.36 & 13:11.2 & 0.477 & 21.97 &&  743 & 36:49.37 & 13:11.25 & 0.45 & 22.12 &\phn\phn    $-0.02  $ \\
 36:49.43 & 13:16.5 & 0.271 & 23.63 &&  759 & 36:49.44 & 13:16.65 & 1.24 & 23.23 &\phn\phn\phs$ 0.76^*$ \\
 36:49.43 & 13:46.8 & 0.089 & 17.97 &&  914 & 36:49.44 & 13:46.91 & 0.06 & 18.16 &\phn\phn    $-0.03  $ \\
 36:49.49 & 14:06.6 & 0.752 & 21.95 &&  989 & 36:49.50 & 14:06.75 & 0.91 & 21.80 &\phn\phn\phs$ 0.09  $ \\
 36:49.57 & 12:20.0 & 0.961 & 24.40 &&  319 & 36:49.51 & 12:20.10 & 0.93 & 24.41 &\phn\phn    $-0.02  $ \\
 36:49.63 & 12:57.6 & 0.475 & 21.91 &&  655 & 36:49.63 & 12:57.57 & 0.46 & 21.99 &\phn\phn    $-0.01  $ \\
 36:49.70 & 13:13.0 & 0.475 & 21.46 &&  746 & 36:49.71 & 13:13.04 & 0.54 & 21.45 &\phn\phn\phs$ 0.04  $ \\
 36:49.72 & 14:14.9 & 1.980 & 23.18 && 1016 & 36:49.82 & 14:14.99 & 1.64 & 23.40 &\phn\phn    $-0.11  $ \\
 36:49.80 & 12:48.8 & 3.233 & 25.13 &&  568 & 36:49.81 & 12:48.79 & 3.50 & 25.18 &\phn\phn\phs$ 0.06  $ \\
 36:49.86 & 12:42.3 & 0.751 & 24.38 &&  506 & 36:49.86 & 12:42.20 & 0.96 & 24.47 &\phn\phn\phs$ 0.12  $ \\
 36:49.95 & 12:25.9 & 1.205 & 23.71 &&  353 & 36:49.94 & 12:25.44 & 1.07 & 24.32 &\phn\phn    $-0.06^+$ \\
 36:50.09 & 14:01.0 & 2.237 & 24.55 &&  960 & 36:50.10 & 14:01.11 & 2.46 & 24.58 &\phn\phn\phs$ 0.07  $ \\
 36:50.15 & 12:16.9 & 0.905 & 23.06 &&  273 & 36:50.16 & 12:16.99 & 0.69 & 22.47 &\phn\phn    $-0.11  $ \\
 36:50.19 & 12:39.8 & 0.474 & 20.43 &&  477 & 36:50.21 & 12:39.74 & 0.42 & 20.67 &\phn\phn    $-0.04  $ \\
 36:50.20 & 13:41.7 & 1.249 & 24.10 &&  883 & 36:50.19 & 13:41.84 & 1.39 & 24.87 &\phn\phn\phs$ 0.06  $ \\
 36:50.26 & 12:45.7 & 0.680 & 21.74 &&  524 & 36:50.26 & 12:45.78 & 0.58 & 21.45 &\phn\phn    $-0.06  $ \\
 36:50.34 & 14:18.5 & 0.819 & 23.41 && 1028 & 36:50.36 & 14:18.65 & 0.85 & 23.09 &\phn\phn\phs$ 0.02  $ \\
 36:50.48 & 13:16.1 & 0.851 & 23.07 &&  749 & 36:50.47 & 13:16.12 & 0.90 & 22.60 &\phn\phn\phs$ 0.03  $ \\
 36:50.82 & 12:55.8 & 0.321 & 22.27 &&  611 & 36:50.82 & 12:55.89 & 0.41 & 22.55 &\phn\phn\phs$ 0.07  $ \\
 36:50.83 & 12:51.5 & 0.485 & 23.15 &&  563 & 36:50.84 & 12:51.47 & 0.56 & 23.16 &\phn\phn\phs$ 0.05  $ \\
 36:51.04 & 13:20.6 & 0.199 & 19.38 &&  757 & 36:51.05 & 13:20.69 & 0.11 & 19.60 &\phn\phn    $-0.07  $ \\
 36:51.17 & 13:48.7 & 3.162 & 25.21 &&  897 & 36:51.20 & 13:48.77 & 2.94 & 25.22 &\phn\phn    $-0.05  $ \\
 36:51.37 & 14:20.9 & 0.439 & 23.22 && 1022 & 36:51.39 & 14:20.91 & 0.45 & 23.31 &\phn\phn\phs$ 0.01  $ \\
 36:51.40 & 13:00.6 & 0.090 & 22.89 &&  637 & 36:51.42 & 13:00.63 & 0.07 & 22.76 &\phn\phn    $-0.02  $ \\
 36:51.69 & 12:20.2 & 0.401 & 21.45 &&  257 & 36:51.71 & 12:20.17 & 0.43 & 21.56 &\phn\phn\phs$ 0.02  $ \\
 36:51.77 & 13:53.7 & 0.557 & 21.08 &&  912 & 36:51.79 & 13:53.81 & 0.57 & 21.10 &\phn\phn\phs$ 0.01  $ \\
 36:51.96 & 13:32.1 & 1.087 & 23.59 &&  801 & 36:51.97 & 13:32.17 & 0.97 & 23.12 &\phn\phn    $-0.06  $ \\
 36:51.96 & 14:00.7 & 0.559 & 23.03 &&  938 & 36:51.98 & 14:00.82 & 0.53 & 23.03 &\phn\phn    $-0.02  $ \\
 36:51.99 & 12:09.6 & 0.458 & 22.75 &&  138 & 36:52.01 & 12:09.67 & 0.53 & 22.90 &\phn\phn\phs$ 0.05  $ \\
 36:52.40 & 13:37.7 & 3.430 & 25.08 &&  824 & 36:52.41 & 13:37.75 & 3.75 & 24.79 &\phn\phn\phs$ 0.07  $ \\
 36:52.66 & 12:19.7 & 0.401 & 23.11 &&  220 & 36:52.67 & 12:19.69 & 0.51 & 23.20 &\phn\phn\phs$ 0.08  $ \\
 36:52.72 & 13:54.7 & 1.355 & 21.85 &&  904 & 36:52.74 & 13:54.77 & 1.43 & 21.98 &\phn\phn\phs$ 0.03  $ \\
 36:52.73 & 13:39.1 & 3.369 & 25.03 &&  825 & 36:52.75 & 13:39.07 & 3.49 & 25.07 &\phn\phn\phs$ 0.03  $ \\
 36:52.84 & 14:04.8 & 0.498 & 23.45 &&  941 & 36:52.87 & 14:04.86 & 0.63 & 23.27 &\phn\phn\phs$ 0.09  $ \\
 36:52.86 & 14:08.2 & 3.367 & 24.49 &&  957 & 36:52.98 & 14:08.47 & 3.48 & 26.85 &\phn\phn\phs$ 0.03^+$ \\
 36:53.16 & 13:22.6 & 2.489 & 24.53 &&  742 & 36:53.18 & 13:22.72 & 2.85 & 24.83 &\phn\phn\phs$ 0.10  $ \\
 36:53.41 & 13:29.3 & 2.991 & 24.64 &&  762 & 36:53.43 & 13:29.41 & 3.49 & 24.60 &\phn\phn\phs$ 0.13  $ \\
 36:53.43 & 12:34.3 & 0.560 & 22.78 &&  335 & 36:53.44 & 12:34.26 & 0.52 & 22.81 &\phn\phn    $-0.03  $ \\
 36:53.59 & 14:10.1 & 3.181 & 24.78 &&  955 & 36:53.59 & 14:10.15 & 3.72 & 24.55 &\phn\phn\phs$ 0.13  $ \\
 36:53.65 & 14:17.6 & 0.517 & 23.36 &&  979 & 36:53.65 & 14:17.62 & 0.45 & 23.45 &\phn\phn    $-0.04  $ \\
 36:53.88 & 12:54.0 & 0.642 & 20.95 &&  500 & 36:53.90 & 12:54.00 & 0.65 & 20.89 &\phn\phn\phs$ 0.00  $ \\
 36:54.07 & 13:54.2 & 0.851 & 22.72 &&  884 & 36:54.09 & 13:54.34 & 1.00 & 22.44 &\phn\phn\phs$ 0.08  $ \\
 36:54.59 & 13:41.3 & 2.419 & 25.20 &&  806 & 36:54.62 & 13:41.31 & 2.40 & 25.27 &\phn\phn    $-0.01  $ \\
 36:54.70 & 13:14.8 & 2.232 & 24.15 &&  670 & 36:54.72 & 13:14.77 & 2.41 & 24.29 &\phn\phn\phs$ 0.06  $ \\
 36:54.96 & 13:14.8 & 0.511 & 23.81 &&  662 & 36:55.00 & 13:14.78 & 0.55 & 23.81 &\phn\phn\phs$ 0.03  $ \\
 36:55.06 & 13:47.3 & 2.233 & 24.23 &&  831 & 36:55.06 & 13:47.06 & 2.40 & 24.52 &\phn\phn\phs$ 0.05  $ \\
 36:55.11 & 13:11.3 & 0.321 & 23.58 &&  631 & 36:55.14 & 13:11.38 & 0.27 & 23.61 &\phn\phn    $-0.04  $ \\
 36:55.14 & 13:03.7 & 0.952 & 24.29 &&  549 & 36:55.15 & 13:03.61 & 0.73 & 23.59 &\phn\phn    $-0.11  $ \\
 36:55.39 & 13:11.0 & 0.968 & 22.86 &&  619 & 36:55.45 & 13:11.19 & 0.87 & 22.16 &\phn\phn    $-0.05  $ \\
 36:55.49 & 14:02.6 & 0.564 & 23.08 &&  899 & 36:55.50 & 14:02.69 & 0.68 & 22.94 &\phn\phn\phs$ 0.07  $ \\
 36:55.51 & 13:53.3 & 1.147 & 22.85 &&  858 & 36:55.52 & 13:53.41 & 1.05 & 22.59 &\phn\phn    $-0.05  $ \\
 36:55.55 & 13:59.8 & 0.559 & 23.74 &&  888 & 36:55.58 & 13:59.88 & 0.64 & 23.70 &\phn\phn\phs$ 0.05  $ \\
 36:55.59 & 12:46.2 & 0.790 & 23.08 &&  381 & 36:55.57 & 12:45.48 & 0.90 & 21.93 &\phn\phn\phs$ 0.06  $ \\
 36:55.59 & 12:49.3 & 0.950 & 23.53 &&  409 & 36:55.62 & 12:49.27 & 0.94 & 22.99 &\phn\phn    $-0.01  $ \\
 36:56.08 & 12:44.7 & 4.022 & 24.94 &&  359 & 36:56.12 & 12:44.68 & 3.85 & 25.48 &\phn\phn    $-0.03  $ \\
 36:56.10 & 13:29.6 & 0.271 & 23.80 &&  734 & 36:56.12 & 13:29.66 & 1.07 & 23.39 &\phn\phn\phs$ 0.63^*$ \\
 36:56.39 & 12:09.3 & 0.321 & 23.22 &&   11 & 36:56.42 & 12:09.23 & 0.43 & 23.27 &\phn\phn\phs$ 0.08  $ \\
 36:56.61 & 12:20.1 & 0.930 & 23.15 &&  104 & 36:56.65 & 12:20.14 & 0.76 & 22.57 &\phn\phn    $-0.09  $ \\
 36:56.61 & 12:52.7 & 1.231 & 23.64 &&  416 & 36:56.59 & 12:52.69 & 0.96 & 23.95 &\phn\phn    $-0.12  $ \\
 36:56.62 & 12:45.5 & 0.518 & 20.06 &&  345 & 36:56.64 & 12:45.32 & 0.69 & 20.00 &\phn\phn\phs$ 0.11  $ \\
 36:56.89 & 13:01.5 & 0.474 & 23.69 &&  486 & 36:56.92 & 13:01.58 & 1.27 & 23.03 &\phn\phn\phs$ 0.54^*$ \\
 36:56.90 & 12:58.0 & 0.520 & 23.84 &&  454 & 36:56.93 & 12:58.19 & 0.58 & 23.59 &\phn\phn\phs$ 0.04  $ \\
 36:57.18 & 12:25.9 & 0.561 & 22.36 &&  144 & 36:57.21 & 12:25.81 & 0.63 & 22.39 &\phn\phn\phs$ 0.04  $ \\
 36:57.27 & 12:59.5 & 0.475 & 21.07 &&  458 & 36:57.31 & 12:59.65 & 0.49 & 21.32 &\phn\phn\phs$ 0.01  $ \\
 36:57.49 & 12:11.0 & 0.665 & 21.10 &&    1 & 36:57.47 & 12:10.56 & 0.71 & 21.20 &\phn\phn\phs$ 0.03  $ \\
 36:57.51 & 12:12.1 & 0.561 & 22.62 &&    6 & 36:57.44 & 12:11.88 & 0.71 & 23.28 &\phn\phn\phs$ 0.10^+$ \\
 36:57.69 & 13:15.3 & 0.952 & 22.94 &&  599 & 36:57.71 & 13:15.19 & 0.86 & 23.02 &\phn\phn    $-0.05  $ \\
 36:58.04 & 13:00.4 & 0.320 & 22.04 &&  445 & 36:58.07 & 13:00.40 & 0.41 & 22.37 &\phn\phn\phs$ 0.07  $ \\
 36:58.35 & 12:14.1 & 1.020 & 23.79 &&    4 & 36:58.30 & 12:14.12 & 1.05 & 23.29 &\phn\phn\phs$ 0.01  $ \\
 36:58.63 & 12:21.8 & 0.682 & 23.40 &&   54 & 36:58.65 & 12:21.71 & 0.73 & 23.36 &\phn\phn\phs$ 0.03  $ \\
 36:58.73 & 12:52.4 & 0.321 & 20.99 &&  350 & 36:58.76 & 12:52.35 & 0.27 & 21.27 &\phn\phn    $-0.04  $ \\
 36:59.01 & 12:23.7 & 4.050 & 25.33 &&   60 & 36:59.10 & 12:23.59 & 4.10 & 25.28 &\phn\phn\phs$ 0.01  $ \\
 36:59.41 & 12:21.5 & 0.472 & 23.53 &&   33 & 36:59.37 & 12:21.65 & 0.45 & 24.04 &\phn\phn    $-0.01  $ \\
 36:59.90 & 12:19.0 & 5.340 & 26.10 &&    3 & 36:59.79 & 12:18.66 & 5.26 & 25.69 &\phn\phn    $-0.01  $ \\
 37:00.05 & 12:25.3 & 2.050 & 23.83 &&   48 & 37:00.08 & 12:25.23 & 2.30 & 23.77 &\phn\phn\phs$ 0.08  $ \\
 37:00.53 & 12:34.7 & 0.563 & 21.43 &&  125 & 37:00.56 & 12:34.61 & 0.44 & 21.37 &\phn\phn    $-0.08  $ \\
\enddata
\tablenotetext{a}{A cross marks those objects with possible double 
identifications listed in Table 1. An asterisk marks the objects which are 
discussed in more detail in Sections 4 and 5.}
\end{deluxetable}

\clearpage
%
%
\begin{deluxetable}{c c c c c r c c c c}
\tablecolumns{10}
\tabletypesize{\scriptsize}
\tablewidth{0pt}
\tablecaption{Objects in \cite{C00} with no spectroscopic 
redshift\tablenotemark{a}}
\tablehead{
\multicolumn{4}{c}{Cohen \etal 2000} & \colhead{} & \multicolumn{5}{c}{Fern\'andez-Soto \etal 1999} \\
\cline{1-4} \cline{6-10} \\
\colhead{RA($-12$h)}         & \colhead{Dec($-62^{\circ}$)} & 
\colhead{$z_{\rm spec}$}     & \colhead{$R$}                & \colhead{} &
\colhead{ID}                 & \colhead{RA($-12$h)}         & 
\colhead{Dec($-62^{\circ}$)} & \colhead{$z_{\rm phot}$}     & 
\colhead{$AB(8140)$}         }
\startdata
 36:39.8 & 12:29\tablenotemark{b} & --- & 23.2 && 669 & 36:39.77 & 12:28.53 &
0.00 & 23.72 \\
 36:39.8 & 12:29\tablenotemark{b} & --- & 23.2 &&  678 & 36:39.71 & 12:29.60 & 1.14 & 24.27 \\
 36:45.3 & 11:43                  & --- & 24.0 &&   81 & 36:45.29 & 11:42.89 & 0.70 & 23.94 \\
 36:47.2 & 13:42                  & --- & 23.9 &&  917 & 36:47.17 & 13:41.88 & 1.16 & 23.72 \\ 
 36:48.5 & 13:17                  & --- & 23.4 &&  775 & 36:48.47 & 13:16.64 & 0.27 & 23.29 \\
 36:52.6 & 12:02\tablenotemark{c} & --- & 23.4 && --- & 36:52.56 & 12:01.62 & 0.02 & 23.52 \\
 36:53.0 & 13:44\tablenotemark{b} & --- & 24.0 &&  852 & 36:53.02 & 13:44.21 & 2.28 & 24.60 \\
 36:53.0 & 13:44\tablenotemark{b} & --- & 24.0 &&  851 & 36:52.96 & 13:43.97 & 1.79 & 25.17 \\
 36:53.3 & 12:22\tablenotemark{b} & --- & 23.7 &&  237 & 36:53.25 & 12:22.72 & 0.77 & 25.02 \\
 36:53.3 & 12:22\tablenotemark{b} & --- & 23.7 &&  212 & 36:53.43 & 12:21.50 & 0.01 & 23.78 \\ 
\enddata
\tablenotetext{a}{One object in Table 10 in C00 (HDF36378\_1235) does not
enter our definition of the ``HDF proper'' area}
\tablenotetext{b}{Two objects in our catalog correspond to one position as
listed by C00}
\tablenotetext{c}{Not an object in FLY99. See text for explanation}
\end{deluxetable}

\clearpage
%
%
\begin{deluxetable}{c c c c c c c c}
\tabletypesize{\scriptsize}
\tablewidth{0pt}
\tablecaption{Photometry for object HDF36526\_1202\tablenotemark{a}}
\tablehead{\colhead{Magnitude} & \colhead{Flux(Error)} & \colhead{Flux(Error)} & \colhead{Flux(Error)} & \colhead{Flux(Error)} & \colhead{Flux(Error)} & \colhead{Flux(Error)} & \colhead{Flux(Error)} \\
\colhead{$AB(8140)$} & \colhead{F300W}  & \colhead{F450W}  & \colhead{F606W}  & \colhead{F814W} & \colhead{$J$}  & \colhead{$H$}  & \colhead{$K$}}
\startdata
 23.52 & 81.36(10.35) & 384.0(5.2) & 834.4(3.6) & 1429.0(6.0) & 1934.0(177.5) & 1617.0(300.5) & 1739.0(252.9) \\
\enddata
\tablenotetext{a}{All fluxes are in nJy}
\end{deluxetable}

\clearpage
%
%
\begin{deluxetable}{c r c c c c r c c c}
\tablecolumns{10}
\tabletypesize{\scriptsize}
\tablewidth{0pt}
\tablecaption{Statistical properties of the $z_{\rm spec}$--$z_{\rm phot}$ comparison}
\tablehead{
\colhead{Redshift} & \multicolumn{4}{c}{Before corrections} & \colhead{} &
\multicolumn{4}{c}{After corrections} \\
\cline{2-5} \cline{7-10} \\
\colhead{range}    & \colhead{$N_{\rm pts}$} & \colhead{$\langle\Delta z/(1+z)\rangle$} & 
\colhead{$\sigma_{\Delta z/(1+z)}$}          & \colhead{\% discordant}      & 
\colhead{}         & \colhead{$N_{\rm pts}$} & \colhead{$\langle\Delta z/(1+z)\rangle$} & 
\colhead{$\sigma_{\Delta z/(1+z)}$}          & \colhead{\% discordant}} 
\startdata
0.0--1.5 & 113 &\phs0.0010 & 0.063 & 6.2 && 109 &\phs0.0003 & 0.063 & 0.0 \\ 
1.5--4.0 &  27 &\phs0.0160 & 0.076 & 7.4 &&  26 &\phs0.0186 & 0.068 & 0.0 \\
4.0--6.0 &   6 &  $-0.0045$& 0.016 & 0.0 &&   6 &  $-0.0045$& 0.016 & 0.0 \\ 
0.0--6.0 & 146 &\phs0.0035 & 0.065 & 6.2 && 141 &\phs0.0035 & 0.064 & 0.0 \\
\enddata
\end{deluxetable}

\clearpage
%
%
\begin{deluxetable}{c c c c c l}
\tabletypesize{\scriptsize}
\tablewidth{0pt}
\tablecaption{Properties of the objects with discordant redshifts}
\tablehead{\colhead{Name} & \colhead{$z_{\rm spec}$} & \colhead{$R$} & \colhead{Q\tablenotemark{a}} & \colhead{$z_{\rm phot}$} & \colhead{Notes}} 
\startdata
 HDF36386\_1234 & 0.904 & 24.04 & 9 & 0.15 & Inconclusive evidence (confusion with nearby object?) \\
 HDF36396\_1230 & 0.943 & 24.40 & 7 & 3.40 & $z=3.475$ must be the real value \\
 HDF36409\_1205 & 0.882 & 22.94 & 9 & 0.00 & Inconclusive evidence (confusion with nearby object?) \\
 HDF36414\_1143 & 0.548 & 23.51 & 3 & 1.32 & Inconclusive evidence (confusion with nearby object?) \\
 HDF36441\_1410 & 2.267 & 24.26 & 1 & 0.01 & $z=0.065$ must be the real value \\
 HDF36450\_1251\tablenotemark{b}& 2.801 & 24.37 & 3 & 2.23 & F300W excess. $z_{\rm spec}$ unlikely \\
 HDF36478\_1256 & 2.931 & 24.35 & 2 & 0.26 & $z_{\rm spec}$ must be right \\
 HDF36494\_1317 & 0.271 & 23.63 & 3 & 1.24 & $z=1.238$ must be the real value \\
 HDF36561\_1330 & 0.271 & 23.80 & 4 & 1.07 & $z=1.238$ must be the real value \\
 HDF36569\_1302 & 0.474 & 23.69 & 1 & 1.27 & $z_{\rm spec}$ wrong due to nearby object. \\
\enddata
\tablenotetext{a}{Spectroscopic quality flag assigned by C00. See text 
for details}
\tablenotetext{b}{This object is not $>4\sigma$-discordant, but strong
evidence points to the spectroscopic redshift being in error. See text for
details}
\end{deluxetable}

\clearpage
\figcaption{Comparison of spectroscopic and photometric redshifts for the 146
objects in the sample. The large squares mark those points which are
discordant at a level $>4\sigma$.}

\figcaption{Comparison of spectroscopic and photometric redshifts for the 146 
objects in the sample. Symbols are as in Figure 1. The dashed lines mark 
the $1\sigma$ interval and the mean value of $\Delta z /(1+z)$ for the 
sample as a whole.}

\figcaption{Distribution of the residuals from the $z_{\rm spec}$--$z_{\rm 
phot}$ comparison. The dotted line is a gaussian distribution with 
mean=0.0035, $\sigma= 0.065.$}

\figcaption{(a--j)Spectral energy distributions of the different models at
$z_{\rm spec}$ (upper panels) and of our best-fit model at $z_{\rm phot}$
(mid panel, empty circles mark the expected flux for each filter), compared
to the observed photometry (points with error bars). The bottom panel in each
figure shows the likelihood function and the values of $z_{\rm spec}$,
$z_{\rm phot}$, and our best-fit type.}

\figcaption{F814W images of all the objects discussed in detail in Sections 3
and 4. The size of each image is 8x8 arcsec.}

\figcaption{Comparison of spectroscopic and photometric redshifts for the 141
objects in the final sample, after the discordant values have been
individually checked. The large square and two error bars joined by a dotted
line correspond to the $2\sigma$ confidence interval around the $z_{\rm
phot}$ value as calculated for HDF36478\_1256.}

\figcaption{Comparison of spectroscopic and photometric redshifts for the 141
objects in the final sample. Symbols are as in Figure 6. The dashed lines
mark the $1\sigma$ interval and the mean value of $\Delta z /(1+z)$ for the
sample as a whole.}

\begin{figure}[t]
\epsscale{1.0}
\plotone{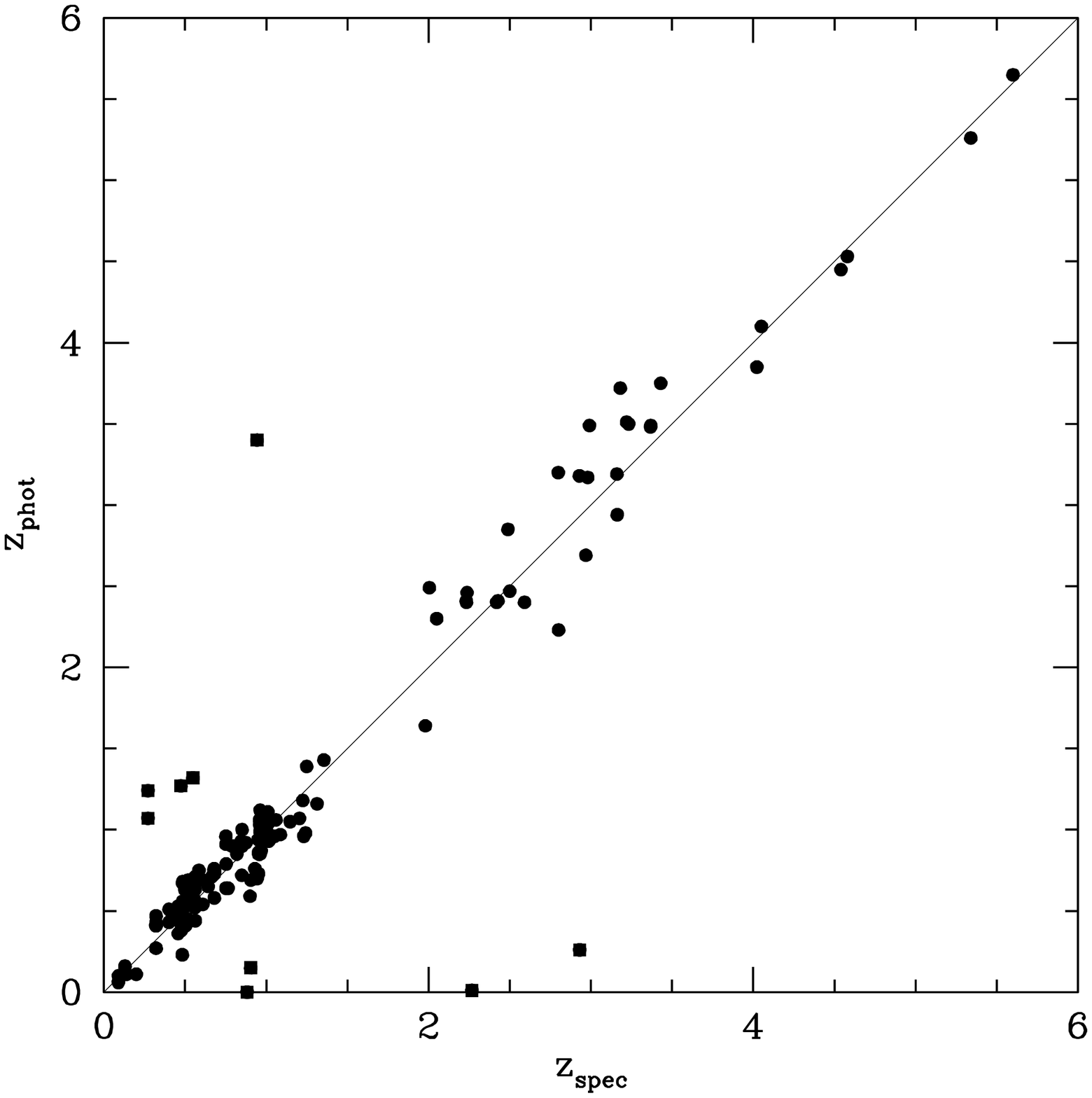}
\end{figure} 

\begin{figure}[t]
\epsscale{1.0}
\plotone{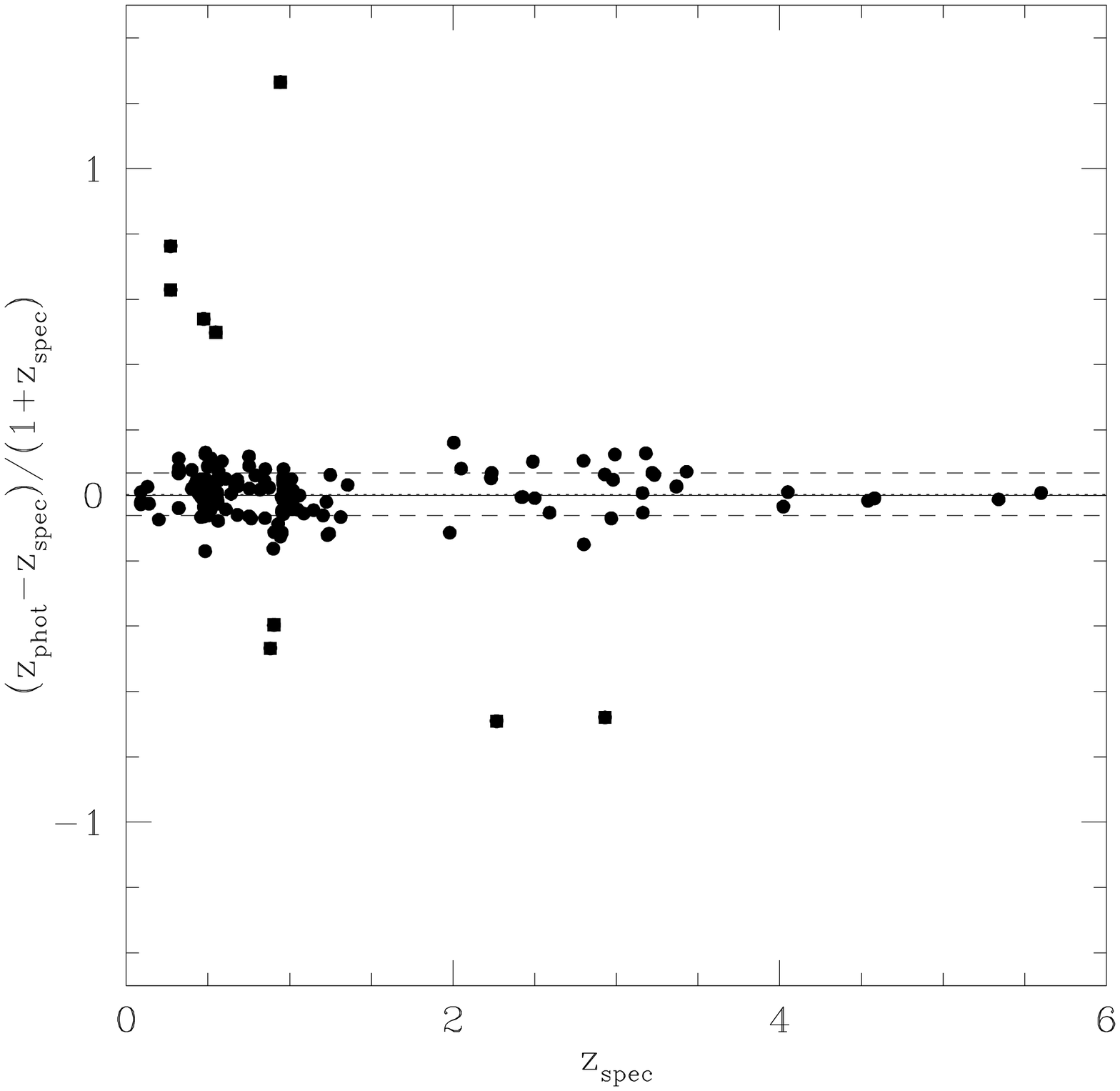}
\end{figure} 

\begin{figure}[t]
\epsscale{1.0}
\plotone{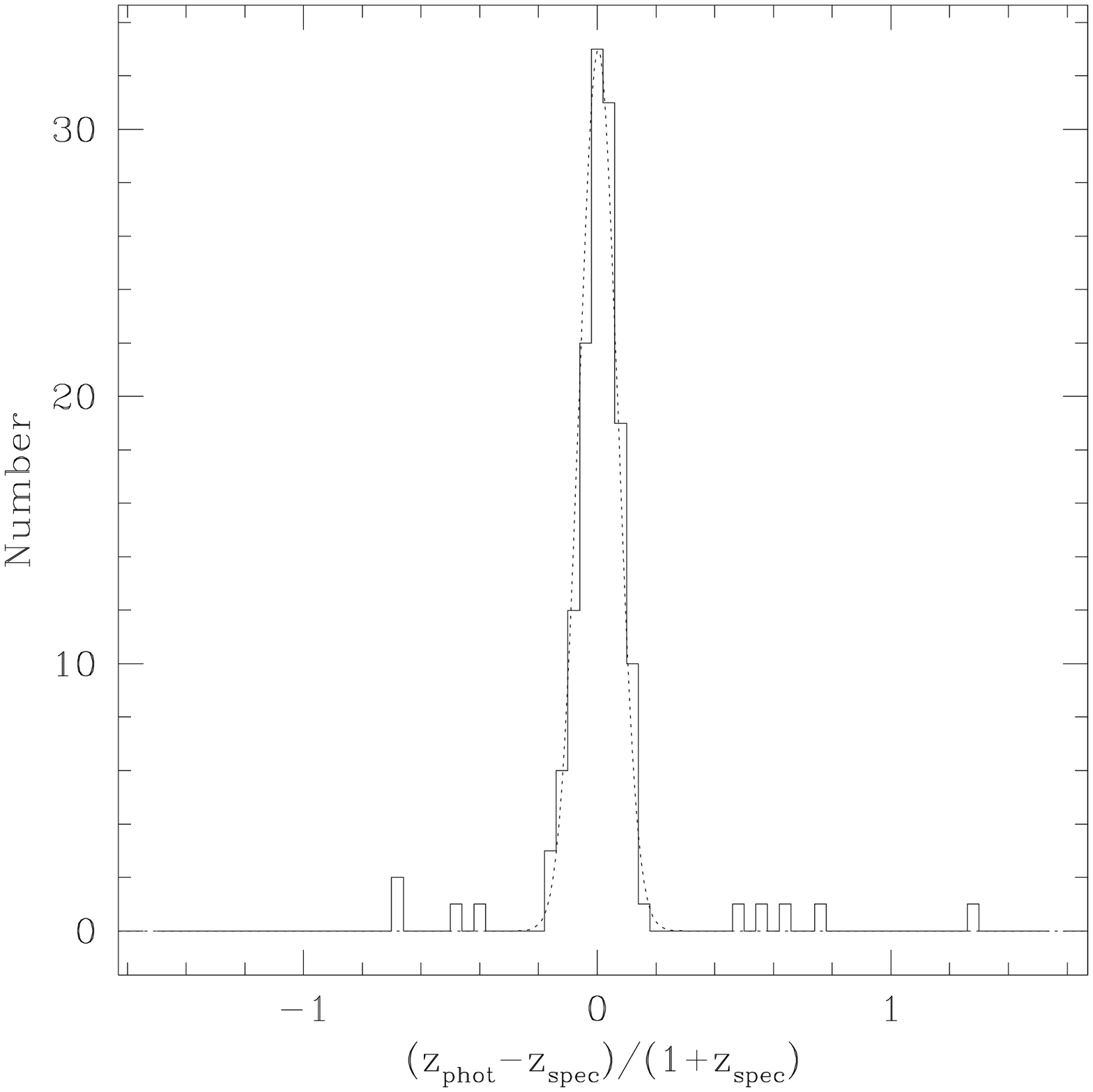}
\end{figure} 

\begin{figure}
\epsscale{1.0}
\plotone{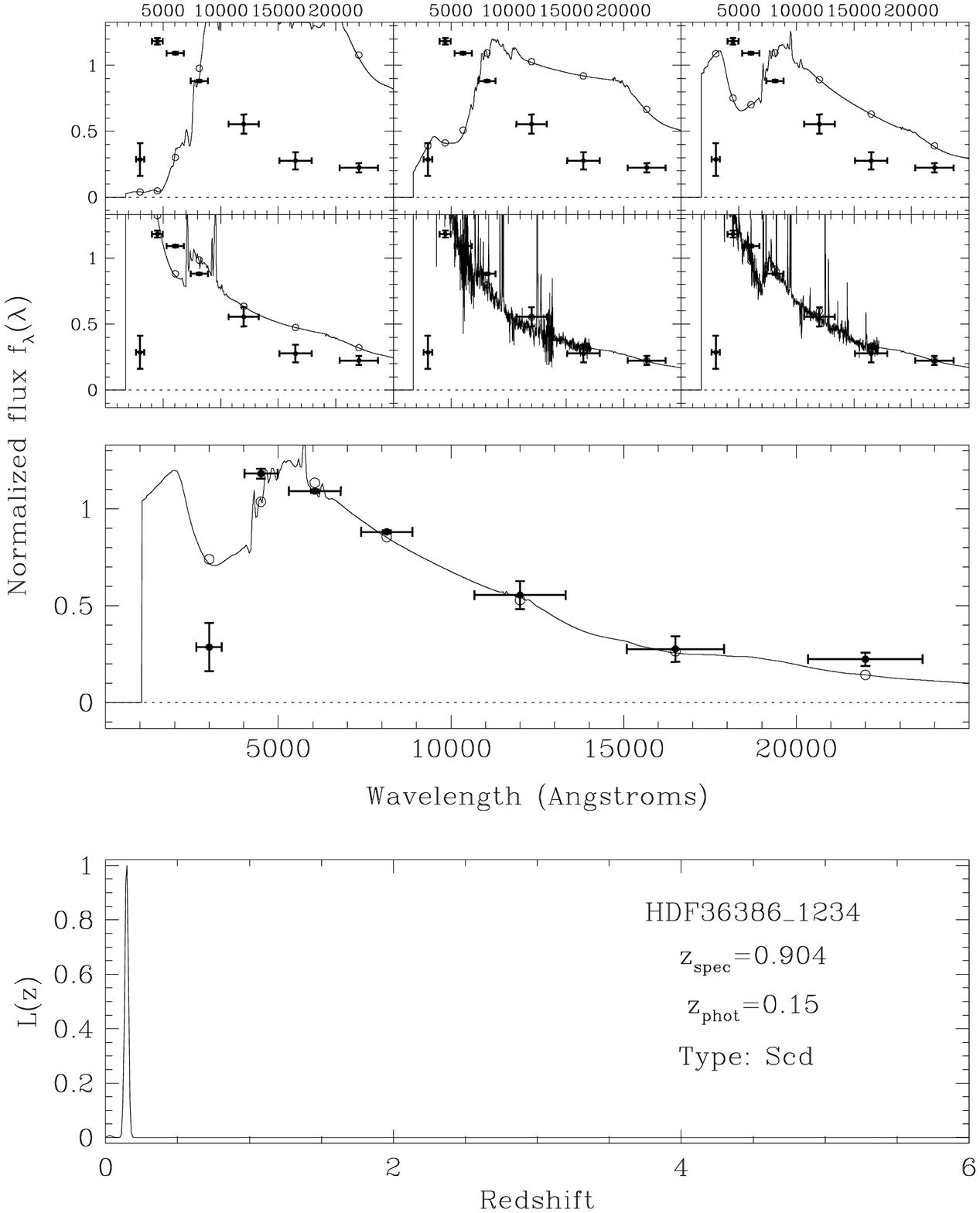}
\end{figure}

\begin{figure}
\epsscale{1.0}
\plotone{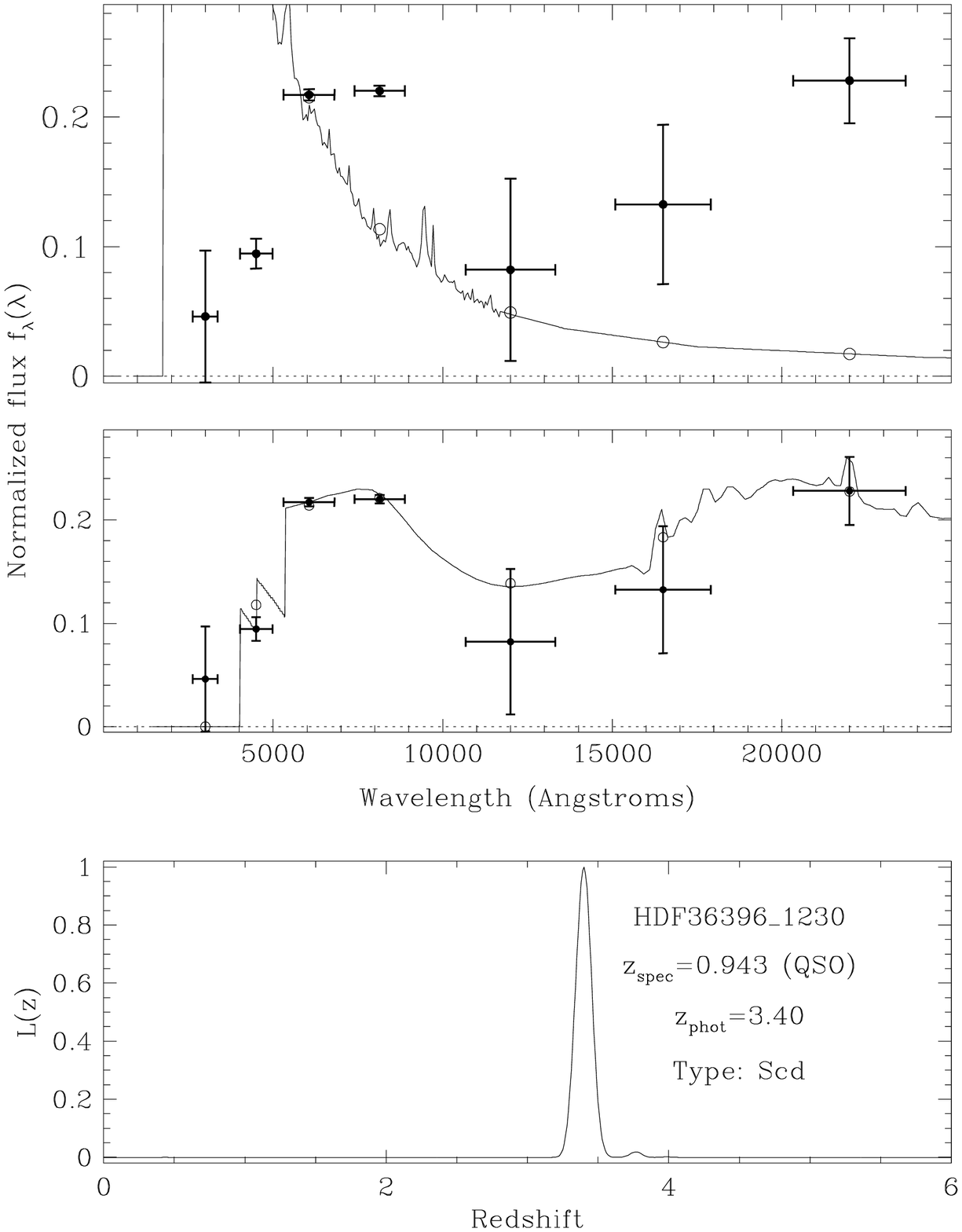}
\end{figure}

\begin{figure}
\epsscale{1.0}
\plotone{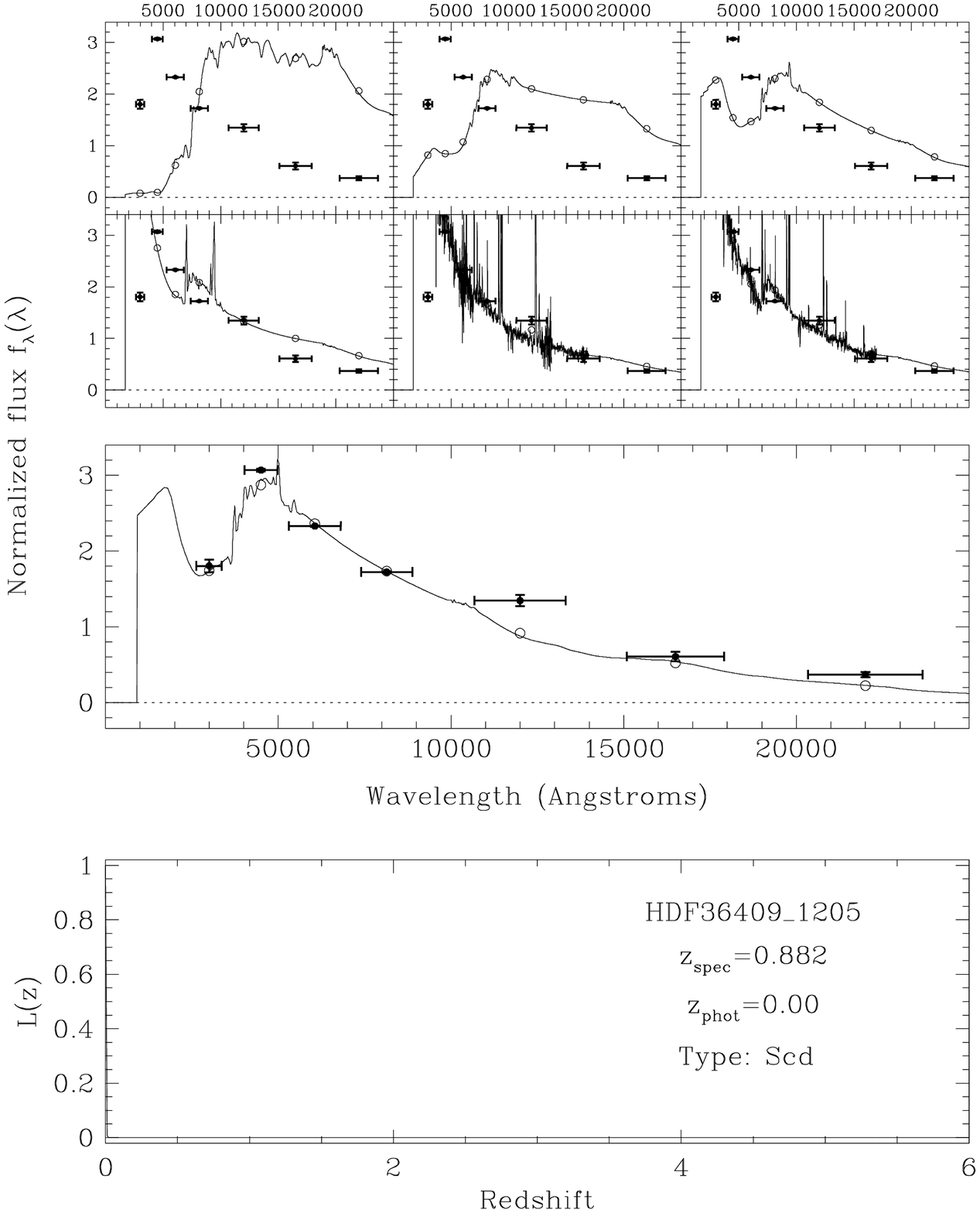}
\end{figure}

\begin{figure}
\epsscale{1.0}
\plotone{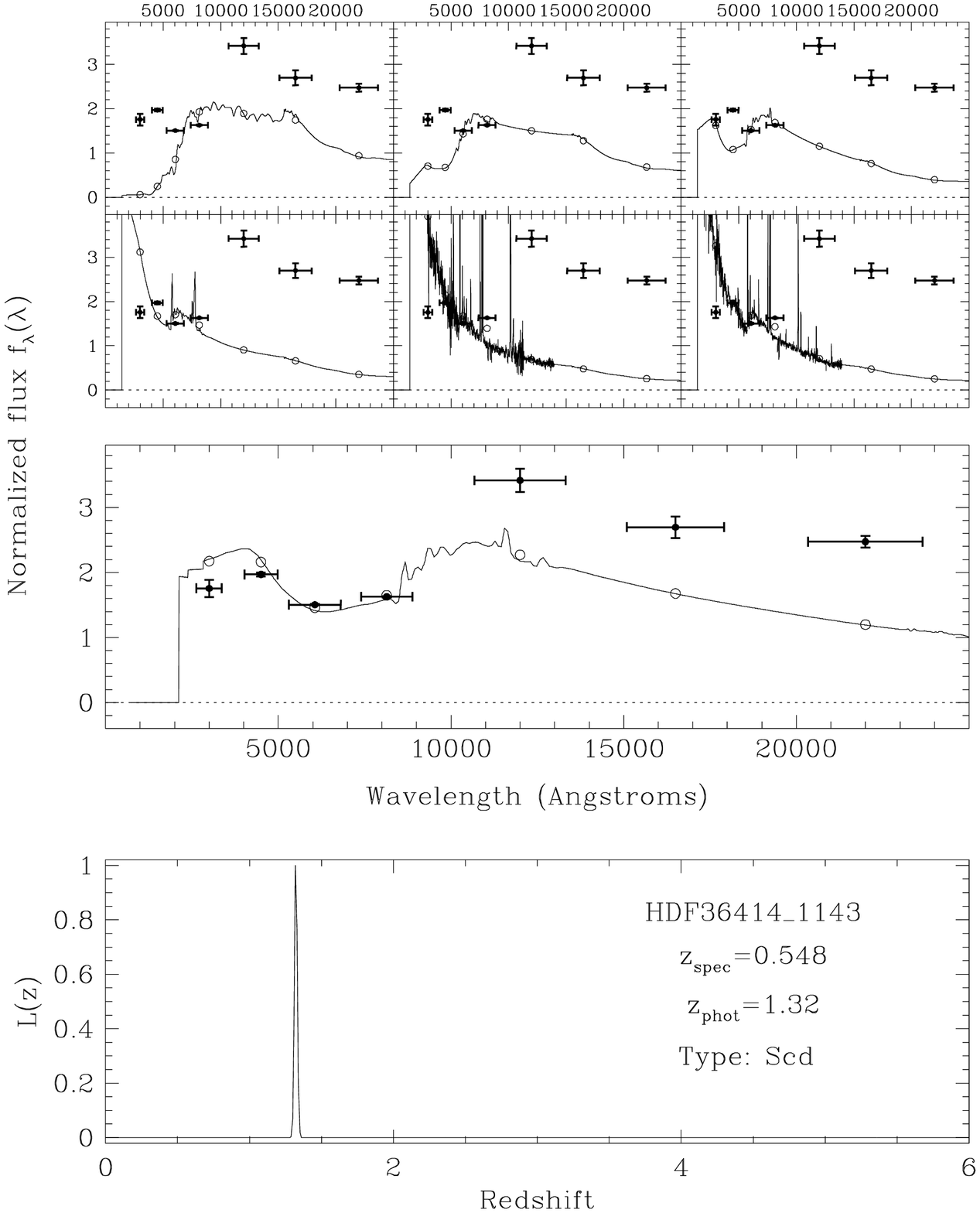}
\end{figure}

\begin{figure}
\epsscale{1.0}
\plotone{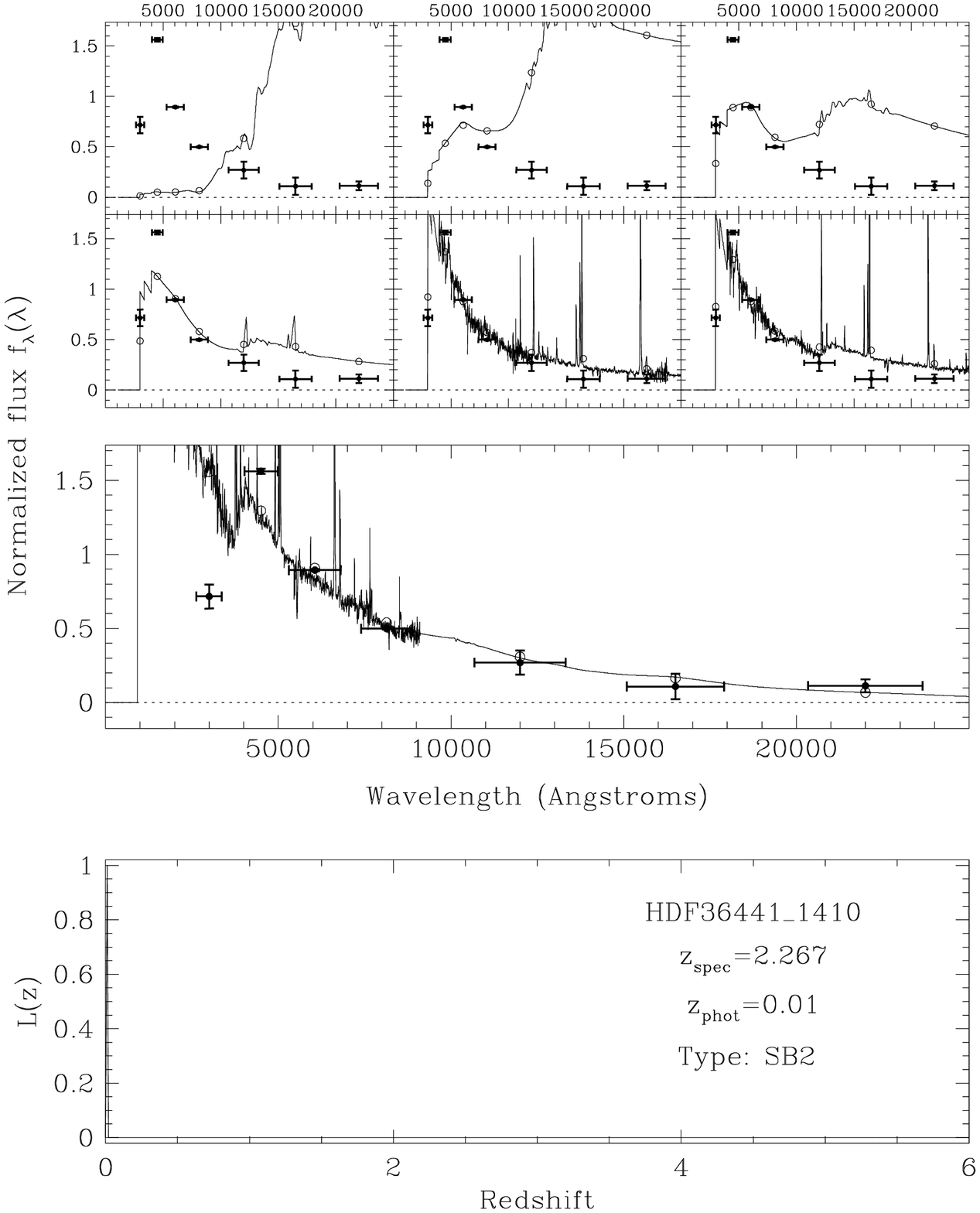}
\end{figure}

\begin{figure}
\epsscale{1.0}
\plotone{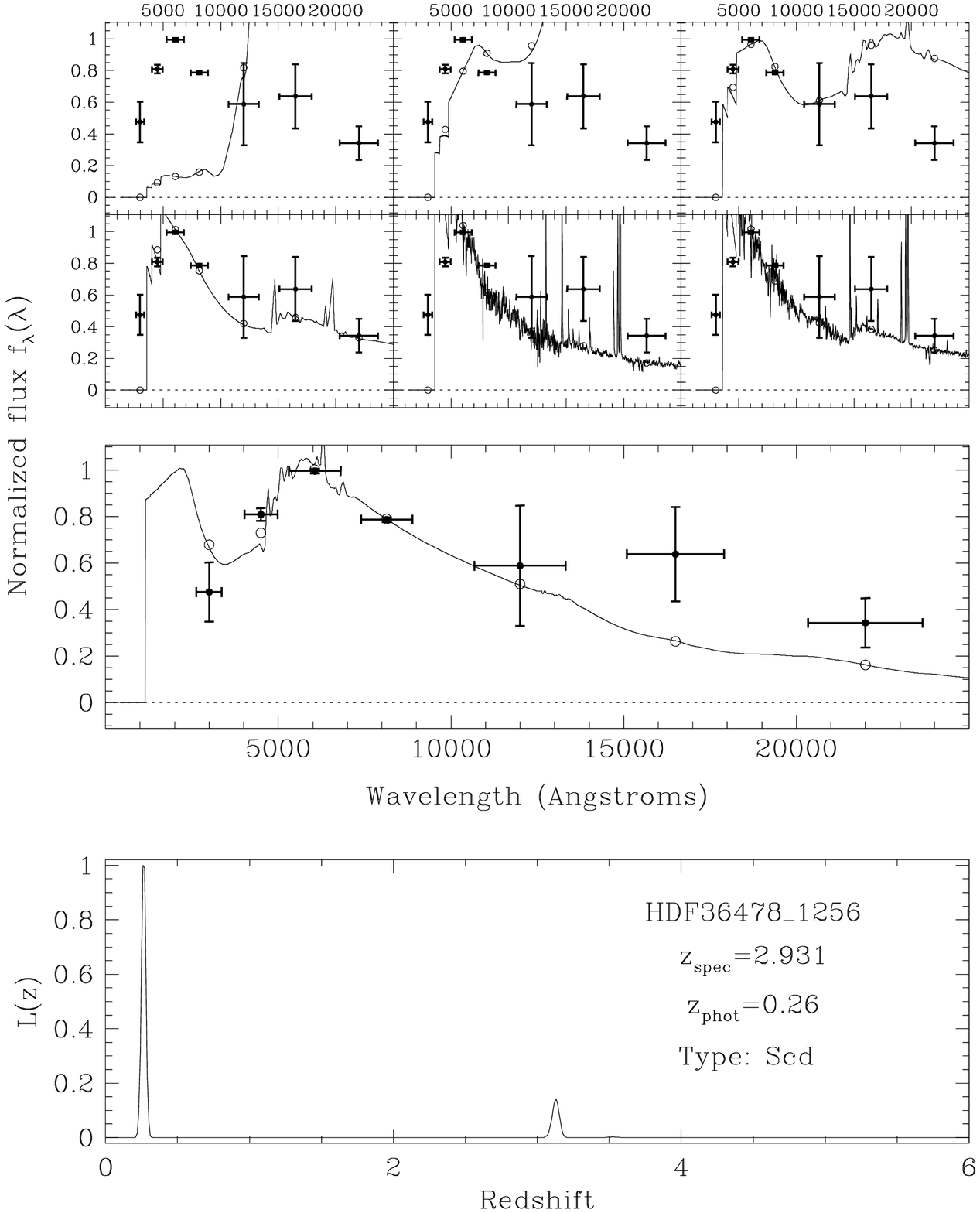}
\end{figure}

\begin{figure}
\epsscale{1.0}
\plotone{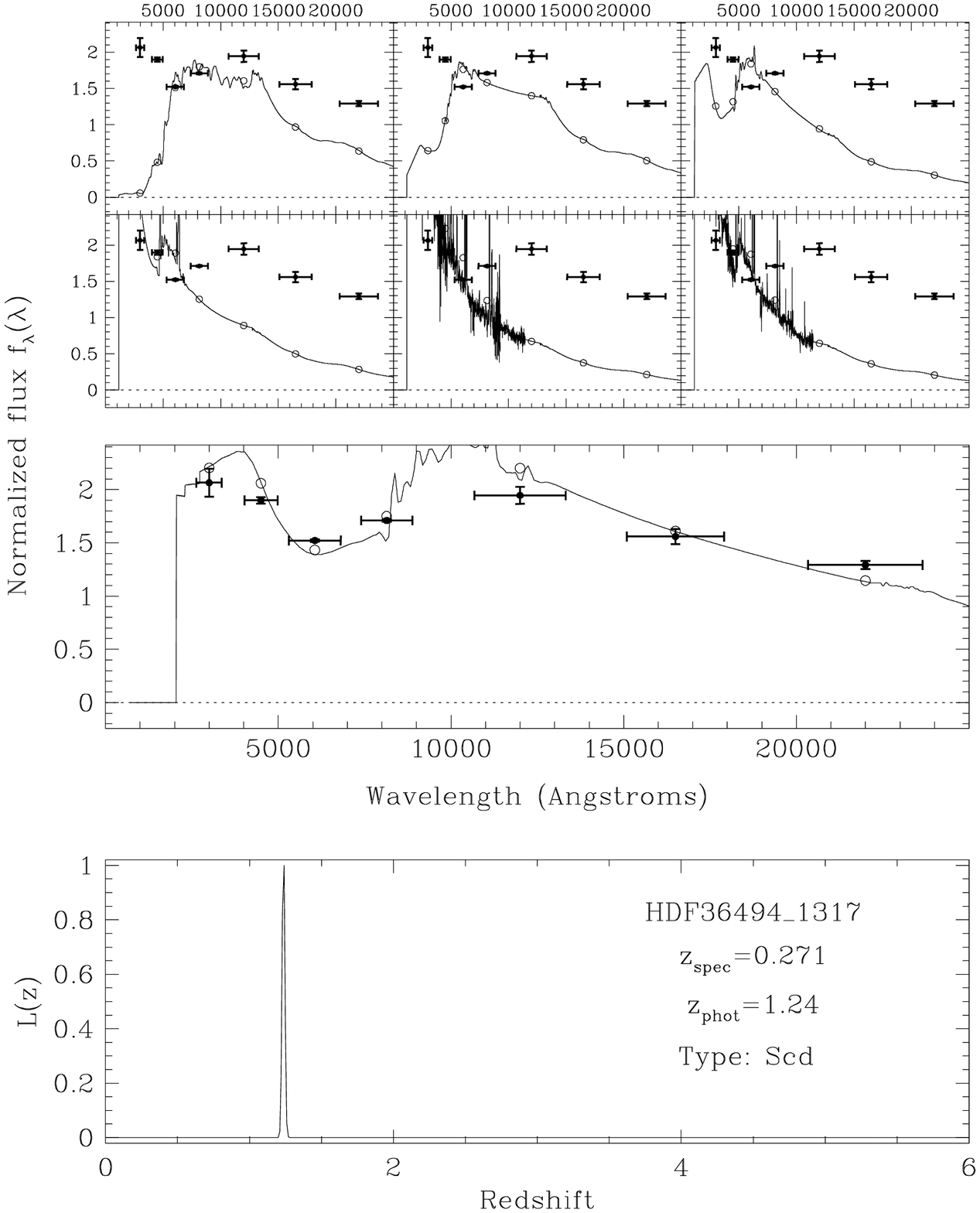}
\end{figure}

\begin{figure}
\epsscale{1.0}
\plotone{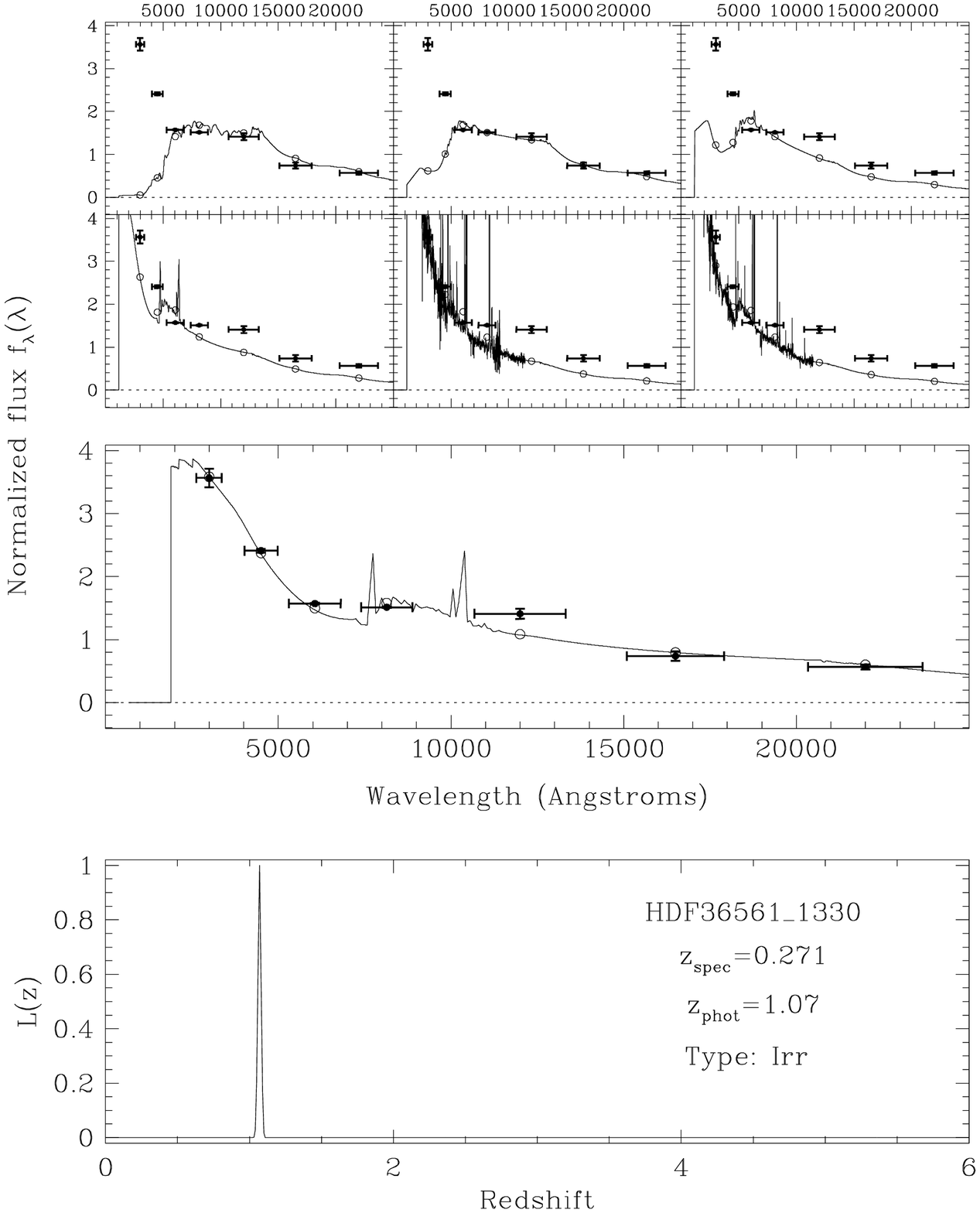}
\end{figure}

\begin{figure}
\epsscale{1.0}
\plotone{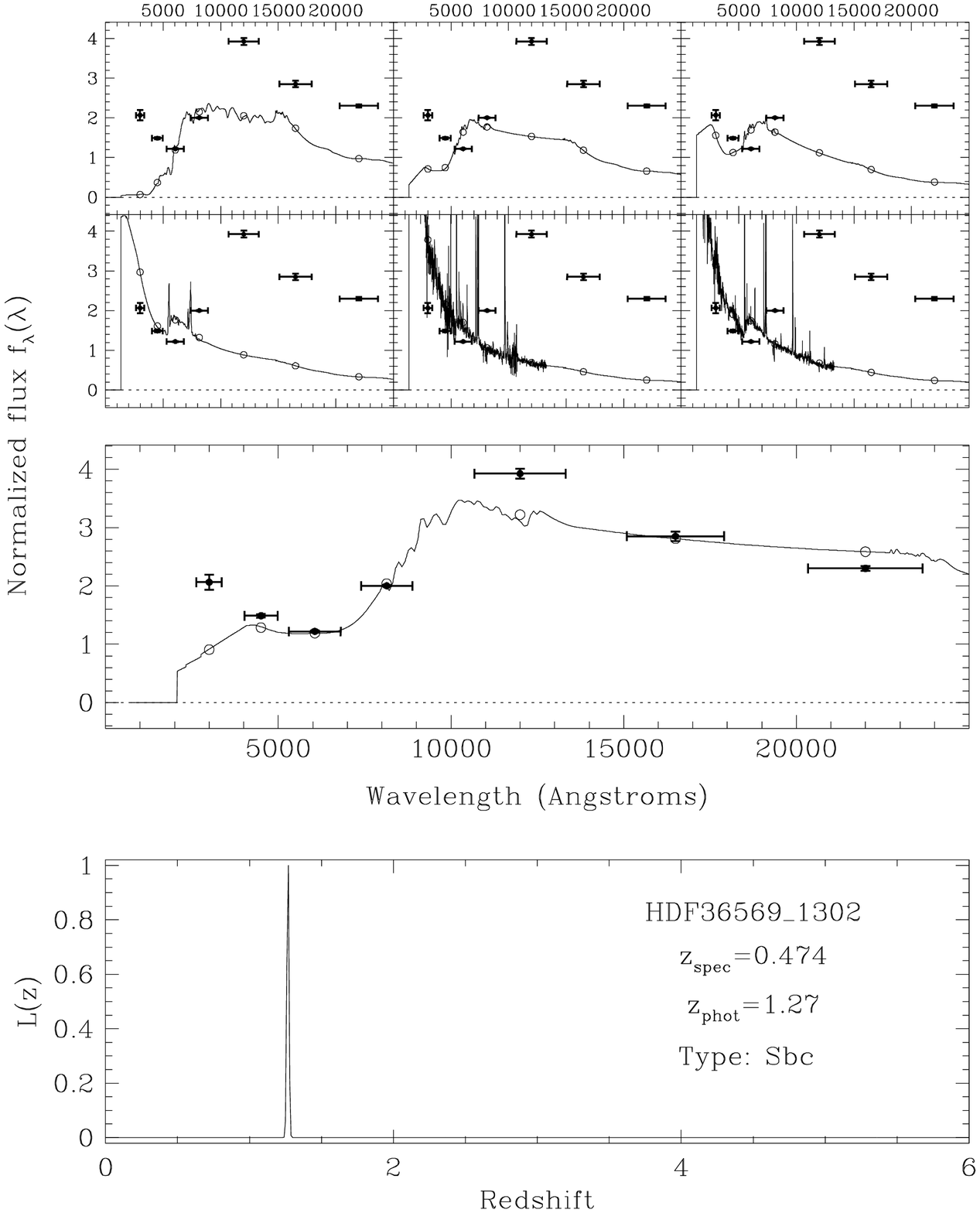}
\end{figure}

\begin{figure}
\epsscale{1.0}
\plotone{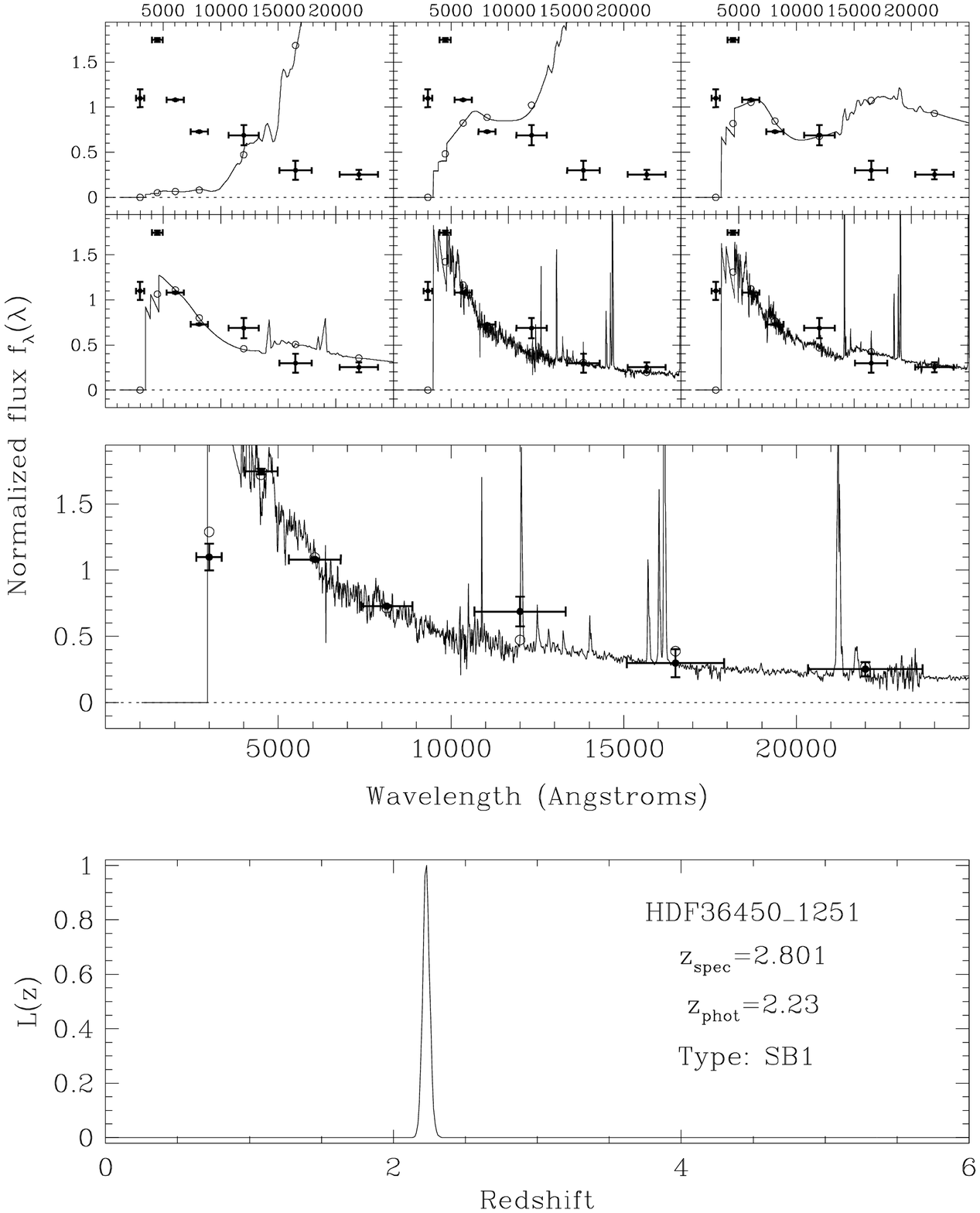}
\end{figure}

\begin{figure}
\epsscale{1.0}
\plotone{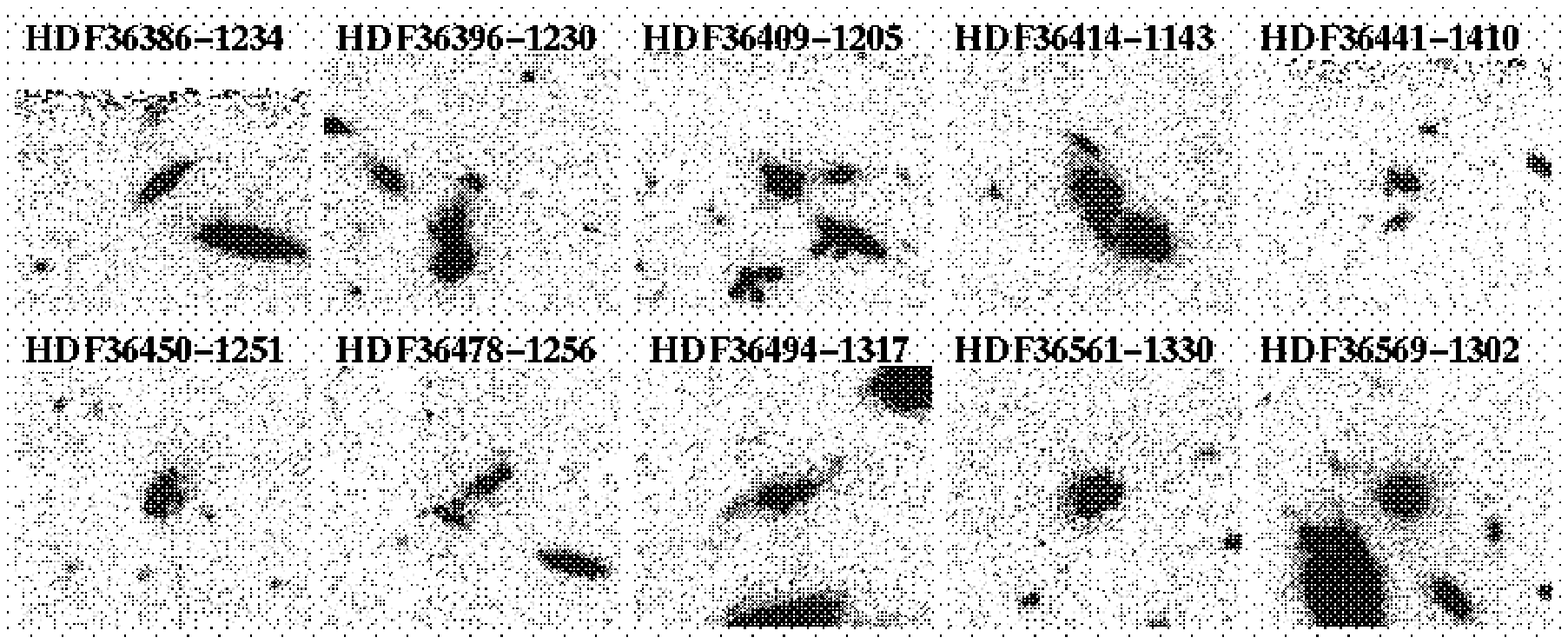}
\end{figure} 

\begin{figure}
\epsscale{1.0}
\plotone{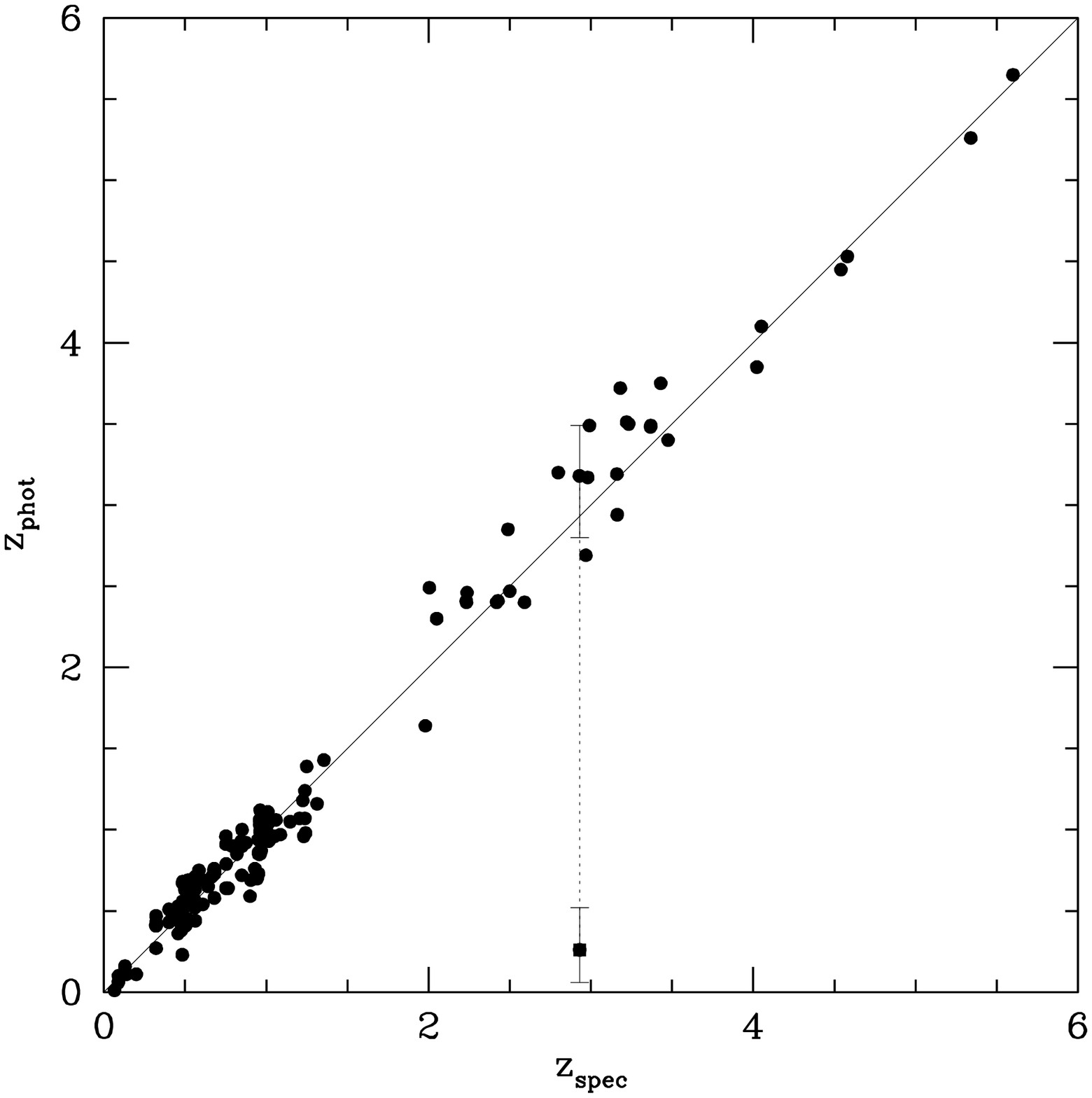}
\end{figure} 

\begin{figure}
\epsscale{1.0}
\plotone{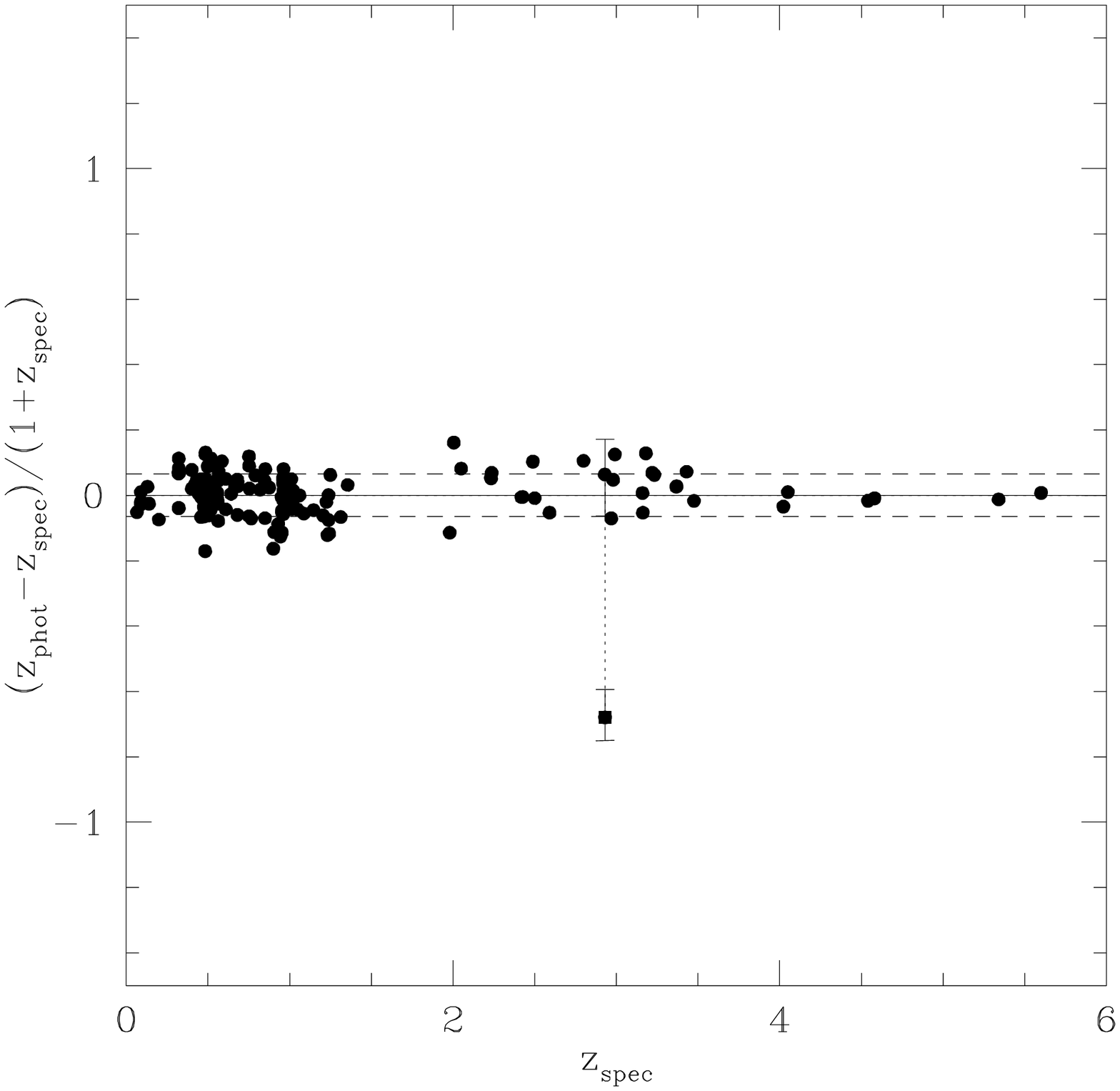}
\end{figure} 

\end{document}